\documentclass[reprint,twocolumn,showpacs,superscriptaddress]{revtex4-1}

\usepackage{epsfig}
\usepackage{amsmath,amssymb}
\usepackage{graphicx}
\usepackage{hyperref}
\hypersetup{colorlinks=true,urlcolor=blue,citecolor=blue,linkcolor=blue,breaklinks=true}
\usepackage[noabbrev]{cleveref}
\usepackage{adjustbox}
\usepackage{floatrow}
\usepackage{footnote}
\usepackage{grffile}
\usepackage{color}
\usepackage{xcolor}
\usepackage{dsfont}
\usepackage{mathtools}
\usepackage{verbatim}
\usepackage{bbold}
\usepackage{hhline}
\usepackage{textcomp}

\allowdisplaybreaks

\newcommand{\cw}{\columnwidth}

\DeclareMathOperator*{\SumInt}{%
\mathchoice%
  {\ooalign{$\displaystyle\sum$\cr\hidewidth$\displaystyle\int$\hidewidth\cr}}
  {\ooalign{\raisebox{.14\height}{\scalebox{.7}{$\textstyle\sum$}}\cr\hidewidth$\textstyle\int$\hidewidth\cr}}
  {\ooalign{\raisebox{.2\height}{\scalebox{.6}{$\scriptstyle\sum$}}\cr$\scriptstyle\int$\cr}}
  {\ooalign{\raisebox{.2\height}{\scalebox{.6}{$\scriptstyle\sum$}}\cr$\scriptstyle\int$\cr}}
}

\begin{document}

\title{Pseudogap opening in the two-dimensional Hubbard model: A functional renormalization group analysis}

\newcommand{\TUVienna}{\affiliation{Institute for Solid State Physics, Vienna University of Technology, 1040 Vienna, Austria}}
\newcommand{\UniTueb}{\affiliation{Institut f\"ur Theoretische Physik and Center for Quantum Science, Universit\"at T\"ubingen, Auf der Morgenstelle 14, 72076 T\"ubingen, Germany}}
\newcommand{\RWTH}{\affiliation{Institute for Theoretical Solid State Physics, RWTH Aachen University, 52056 Aachen, Germany}}
\newcommand{\Jara}{\affiliation{JARA-FIT, RWTH Aachen University, 52056 Aachen, Germany}}
\newcommand{\Julich}{\affiliation{SimLab Quantum Materials, J\"ulich Supercomputing Centre, Forschungszentrum J\"ulich GmbH,
D-52425 J\"ulich, Germany}}

\author{Cornelia Hille}       \UniTueb

\author{Daniel Rohe}  \Julich

\author{Carsten Honerkamp} 	\RWTH\Jara

\author{Sabine Andergassen} 	\UniTueb

\begin{abstract}
Using the recently introduced multiloop extension of the functional renormalization group, we compute the frequency- and momentum-dependent self-energy of the two-dimensional Hubbard model at half filling and weak coupling. We show that, in the truncated-unity approach for the vertex, it is essential to adopt the Schwinger-Dyson form of the self-energy flow equation in order to capture the pseudogap opening. We provide an analytic understanding of the key role played by the flow scheme in correctly accounting for the impact of the antiferromagnetic fluctuations. For the resulting pseudogap, we present a detailed numerical analysis of its evolution with temperature, interaction strength, and loop order.
\end{abstract}

\maketitle

\section{Introduction}

In correlated electrons physics, the term pseudogap is usually associated with a gap-like suppression of the low-energy spectral weight that occurs without a direct connection to a phase transition. While in typical one-dimensional conductors, the pseudogap can be understood as a precursor of density-wave or Peierls ordering \cite{Lee1973,Sadovskii1979,Bartosch1999}, in other two-dimensional (2D) systems, in particular, in hole-doped cuprates, the mechanism of the pseudogap and its connection to symmetry-breaking transitions are less clear. The associated momentum anisotropy in the spectral function of the cuprates, with a pronounced quasiparticle gap at the antinodal point, has been observed in numerous experiments (see, e.g., \cite{Keimer2015,Comin2014}). 
Theoretically, the pseudogap has been identified in the hole-doped 2D Hubbard model with a small next-nearest-neighbor hopping amplitude $t'$ and strong coupling \cite{Huscroft2001,Senechal2004,Kyung2006,Macridin2006,Tremblay2006,Gull2009,Gull2010,Gull2012,Scalapino2012,Gunnarsson2015,Gunnarsson2016}. A momentum-selective gap opening has been observed also at electron doping, for weak to intermediate interaction strengths \cite{Senechal2004,Kyung2003,Kyung2004,Hankevych2006,Wu2017,Gull2013,Vilk1997,Tremblay2006}. In contrast to the pseudogap originating from strong coupling effects, the weak-coupling mechanism is induced by long-range antiferromagnetic (AF) correlations \cite{Vilk1996,Vilk1997,Moukouri2000,Kyung2003,Senechal2004,Kyung2004,Hankevych2006,Schaefer2015,Wu2017}. 
For this reason also the half-filled Hubbard model without next-nearest-neighbor hopping has been considered, with the two-particle self-consistent approach \cite{Vilk1997}, the dynamical cluster approximation \cite{Gull2013}, the dynamical vertex approximation \cite{Schaefer2015}, and recently also with the parquet approximation \cite{Eckhardt2019} and self-energy diagrammatic determinant Monte Carlo \cite{Simkovic2020} (mostly at weak to intermediate coupling). 

To capture pseudogap effects in fRG calculations at weak to intermediate coupling, a proper resolution of the sharp AF peak in the magnetic vertex is essential and requires a fine transfer momentum grid. This can be implemented very efficiently within the advanced variants of the functional renormalization group (fRG), like the recently described truncated-unity fRG \cite{Lichtenstein2017,TagliaviniHille2019,Hille2020} (TU-fRG), where the bosonic transfer momentum dependence can be suitably adapted, while the fermionic momentum dependencies are sufficiently parametrized by fewer form factors. We present here the observation of pseudogap physics within the TU-fRG. Our reasoning leading to this finding also sheds lights on an apparent mystery around two previous advanced fRG studies \cite{Vilardi2017,Vilardi2018}. In the first fRG work \cite{Vilardi2017}, including the full frequency dependence of the two-particle vertex together with a truncation to $s$-wave form factors, the authors did not report a strong momentum dependence in the quasiparticle weight. On the other hand, in the second publication \cite{Vilardi2018}, using an \textit{a posteriori} evaluation of the self-energy with the Schwinger-Dyson equation \cite{Schwinger,Dyson} (SDE), a gap at the antinodal point was observed. 

In previous $N$-patch momentum discretizations, a pseudogap was detected in the quasiparticle weights computed from two-loop contributions to the self-energy. These were obtained either by using the Wick-ordered fRG \cite{Rohe2005} or by inserting the integrated one-loop equation for the vertex into the flow equation for the self-energy \cite{Katanin2004}, avoiding in this way an explicit computation of a frequency dependent two-particle vertex. Other momentum-resolved quantities obtained in $N$-patch fRG were also indicative of a partial loss of the Fermi surface \cite{Honerkamp2001a,Honerkamp2001b}. We note that only more recent studies \cite{Uebelacker2012,Vilardi2017} include the frequency dependence of the vertex, finding however no momentum-dependent gap \cite{Honerkamp2003,Uebelacker2012,Vilardi2017}. 
Mostly, models including a finite next-nearest-neighbor hopping $t'$ have been considered. For those, a gap opening has been observed at hole doping, not too far from Van Hove filling, at the hot spots where the Fermi surface crosses the magnetic Brillouin zone \cite{Rohe2005,Honerkamp2001a}. Non-Fermi liquid behavior of the self-energy has been observed close to the pseudo-critical temperature and in the immediate vicinity of the hot spots. 
Without next-nearest-neighbor hopping, the quasiparticle weight at $T=0$ appears to be strongly renormalized at the antinodal point, and less at the nodal one \cite{Zanchi2001}, with no pronounced gap-opening tendencies \footnote{The flow of the quasi-particle weight \cite{Honerkamp2003} exhibits only a weak quasi-particle weight suppression as well as momentum anisotropy.}. 

In this work, we use a forefront algorithmic implementation of the fRG \cite{TagliaviniHille2019}, which, with respect to previous fRG-based computation schemes, includes also an accurate treatment of the frequency dependence of the two-particle vertex. 
We provide a reasoning for the lack of pseudogap physics in previous, conventional one-loop ($1\ell$) flows of the self-energy and demonstrate that its replacement by the derivative of the SDE yields the expected gap opening \footnote{For related fRG schemes using the SDE, see Refs.~[\onlinecite{Bartosch2009,Veschgini}].}.  
More precisely, the pseudogap is defined as the change of the self-energy frequency dependence from Fermi-liquid-like to insulator-like at temperatures above which (or interaction strengths below which) the fRG flow signals an apparent AF ordering, which we refer to as pseudo-criticality. The term ``pseudo'' indicates that finite-$T$ magnetic ordering should not occur in these 2D models. Recent work using the related parquet approximation \cite{Eckhardt2019} suggests that the fRG may indeed be capable of capturing this Mermin-Wagner-physics.   
On the technical level, consistent with the recently introduced multiloop extension of the fRG \cite{Kugler2018a,Kugler2018b}, the flow equation for the self-energy has to be adapted \cite{Hille2020} according to the SDE in the TU-fRG.
In fact, it is known that the SDE takes into account spin excitations explicitly, including the second order self-energy corrections for an effective model in which the high energy spin fluctuations are integrated out \cite{Vilk1997,Abanov2003,Montiel2017,Wu2017}. By using the Ornstein-Zernike form for the spin susceptibility \cite{Furman1974,Dare1996}, the SDE predicts a spectral gap for momenta close to the hot spots \cite{Vilk1997,Wu2017}. Our modified TU-fRG scheme captures this physics truthfully and is hence capable of resolving the pseudogap.  

The paper is organized as follows:
In Section \ref{sec:modelandmethod} we introduce the Hubbard model and describe the SDE flow scheme employed for the computation of the self-energy, which in the TU-fRG framework correctly accounts for the form-factor projections in the different channels.
We present our results in Section \ref{sec:1l_vs_SDE}, showing that the long-range AF fluctuations lead to a gap opening in the SDE approach, in contrast to the conventional $1\ell$ flow of the self-energy. 
Close to the AF pseudo-transition, we find a momentum-dependent gap which is maximal at the antinodal point and, depending on the parameter regimes, vanishes in the antinodal one. A qualitative explanation of this method dependence is given in \ref{ssec:SEinTUfRG}. We conclude with a summary and an outlook in Section \ref{sec:conclusion}.

\section{Model and method}
\label{sec:modelandmethod}

\subsection{2D Hubbard model}

We present here results for a prototypical model of correlated fermions, the 2D Hubbard model. In standard second quantization the Hubbard Hamiltonian reads 
\begin{equation}
H=-t\sum_{\langle ij\rangle \sigma }\hat{c}_{i\sigma }^{\dagger }\hat{c}_{j\sigma}+U\sum_{i}\hat{n}_{i\uparrow }\hat{n}_{i\downarrow } - \mu \sum_{i,\sigma}\hat{n}_{i\sigma}\label{H}\;,
\end{equation}
where $t$ denotes the nearest-neighbor hopping amplitude on a square lattice and $U$ the local Coulomb repulsion. In the following, we define our energies in terms of $t\equiv 1$ and restrict our analysis to half filling $\langle \hat{n} \rangle=1$. This is achieved by an implicit shift of the Hartree part by $U\langle \hat{n}_\sigma \rangle$ and setting the chemical potential to $\mu=0$. In this case, the momentum transfer of $(\pi, \pm \pi)$ corresponds to perfect AF nesting on the square-shaped Fermi surface.

\subsection{Functional renormalization group}
\label{ssec:fRG}

The characteristic scale-dependent behavior of numerous strongly correlated electron systems can be treated in a flexible and unbiased way by the fRG, see Refs.~\cite{Salmhofer1999,Berges2002,Kopietz2010,Metzner2012} for a review. Its starting point is an exact functional flow equation, which yields the gradual evolution from a microscopic model action to the final effective action as a function of a flow scheme dependent energy scale. By expanding in powers of the fields one obtains an exact hierarchy of flow equations for vertex functions, which in practical implementations is restricted to the one- and two-particle vertex.
Neglecting the renormalization of three- and higher order particle vertices yields approximate $1\ell$ flow equations for the self-energy and two-particle vertex.
The underlying approximations are devised for the weak to moderate coupling regime, where forefront algorithmic advancements brought the fRG for interacting fermions on 2D lattices to a quantitatively reliable level \cite{TagliaviniHille2019,Hille2020}. 

The substantial improvement with respect to previous fRG-based computation schemes relies on an efficient parametrization of the two-particle vertex which takes into account \emph{both} the momentum and frequency dependence.
Assuming SU(2) spin-rotation symmetry, the one-particle irreducible vertex $V_{\sigma_1,\sigma_2,\sigma_3,\sigma_4}(k_1,k_2,k_3,k_4)$ can be expressed by a coupling function depending on three independent generalized momenta $V(k_1,k_2,k_3)$ via \cite{Honerkamp2001a} 
\begin{align}
    V(k_1,k_2,k_3,k_4)_{\sigma_1,\sigma_2,\sigma_3,\sigma_4}=& -\delta_{\sigma_1,\sigma_4}\delta_{\sigma_2,\sigma_3} V(k_1,k_4,k_3)\nonumber \\
    &\hspace{-0cm}+ \delta_{\sigma_1,\sigma_2}\delta_{\sigma_3,\sigma_4}  V(k_1,k_2,k_3) \; ,
\end{align}
with $k_4=k_1+k_3-k_2$. 
Then the coupling function can be channel-decomposed by the parquet decomposition \cite{Karrasch2008}
\begin{align}
    V(k_1,k_2,k_3)=&\,\Lambda_{\mathrm{2PI}}+\Phi_{pp}(k_1+k_3,k_1,k_4) \nonumber \\
    &\hspace{.76cm}+\Phi_{ph}(k_2-k_1,k_1,k_4)\nonumber\\
    &\hspace{.76cm}+\Phi_{\overline{ph}}(k_3-k_2,k_1,k_2)\; ,
\end{align}
where in the parquet approximation, the fully two-particle irreducible vertex is approximated by $\Lambda_{\mathrm{2PI}}=U$, 
and the two-particle reducible contributions $\Phi_{pp/ph/\overline{ph}}$ in the particle-particle, particle-hole, and crossed (or transverse) particle-hole channel are parametrized using a single generalized 'bosonic' transfer momentum/frequency as first argument and two 'fermionic' ones as second and third argument.
In particular, we combine the TU-fRG \cite{Husemann2009,Wang2012,Lichtenstein2017} using a form-factor expansion for the fermionic momentum dependences with the full frequency treatment including the fermionic high-frequency asymptotics \cite{Rohringer2012,Wentzell2016}, 
see Refs.~\cite{TagliaviniHille2019} and~\cite{Hille2020} for the details on the algorithmic implementation.

In addition, we compute the self-energy and its feedback in the fRG flow of the two-particle vertex (see also Refs.~\cite{Eberlein2015} and~\cite{Vilardi2017}).
Instead of the conventional $1\ell$ flow equation for the self-energy, which we recall for completeness
\begin{align}
\label{eq:Sigma_1l}
    \dot{\Sigma}^{\Lambda}(k)=-\sum_p \Big(2 V^{\Lambda}(k,k,p)-V^{\Lambda}(p,k,k) \Big) S^{\Lambda}(p)\;,
\end{align}
we employ here the scale derivative of the SDE \cite{Schwinger,Dyson}. This is inspired by its connection to the multiloop extension of the fRG, which allows to sum up all the diagrams of the parquet approximation with their exact weight \cite{Kugler2018a,Kugler2018b} and hence yields cutoff-independent results \cite{TagliaviniHille2019}. 
Specifically, the multiloop equation for the self-energy flow can be derived from the SDE 
\begin{align}
    \Sigma({\bf k},i\nu)=& \,U\sum_{{\bf k'},i\nu'}G({\bf k'},i\nu') e^{\pm i \nu 0^+} \nonumber \\ 
    &\hspace{-.9cm}-\sum_{{\bf k' q}} \sum_{i\nu' i\omega} 
    V({\bf k},{\bf k'},{\bf k}+{\bf q},i\nu,i\nu',i\nu+i\omega) \nonumber\\
    &\hspace{-.9cm}\times G({\bf k'},i\nu') G({\bf k}+{\bf q},i\nu+i\omega) 
    G({\bf k'}+{\bf q},i\nu'+i\omega)\, U \; ,
\label{eq:SDE}
\end{align}
with $\Sigma(k)=\delta_{\sigma_1,\sigma_2}\Sigma_{\sigma_1,\sigma_2}(k)$, an implicit factor $T$ for each frequency sum, and the normalization of the momentum sum with respect to the first Brillouin zone.
Here we do not aim at a quantitative analysis within the multiloop fRG, for which the comparison to determinant Quantum Monte Carlo data showed that the fRG is remarkably accurate up to moderate interaction strengths \cite{Hille2020}.
We rather address a qualitative description of the spectral properties (with a substantially reduced numerical effort with respect to fully loop converged computations) retaining only the multiloop equation for the self-energy, i.e. the scale derivative of the SDE.  
Differently to the conventional $1\ell$ flow, the scale derivative of the SDE accounts for the form-factor truncation \cite{Hille2020}.
Formally, the derivative with respect to the flow parameter $\Lambda$ yields the conventional $1\ell$ contribution as well as the multiloop corrections. The proof thereof reported in Ref.~\cite{Kugler2018b}
assumes the equivalence between the different channel representations of the SDE.
However, in the TU-fRG only a finite number of form factors is considered in each channel (typically restricted to just a few). In this approximation, the transfer momentum dependence is not fully reconstructed after a translation from one to another, i.e. the projection between different channel representations is affected by information losses.
We cured this by implementing the self-energy flow, considering the most favorable channel representation for each contribution, directly in the derivative of the SDE (see also Ref.~\cite{Hille2020})
\begin{align} 
    \dot{\Sigma}({\bf k},i\nu) =&\, \dot{\Sigma}_{G}(k,i\nu)   
     + \dot{\Sigma}_{GGG}(k,i\nu) \nonumber\\
     &+ \dot{\Sigma}_{pp}(k,i\nu) 
     + \dot{\Sigma}_{ph}(k,i\nu) 
     + \dot{\Sigma}_{\overline{ph}}(k,i\nu)    \label{eq:sigmasplit} \;,
\end{align}
where the dot represents the $\Lambda$-derivative, and 
\begin{widetext}
\begin{align} 
    \dot{\Sigma}_{G}({\bf k},i\nu) =&\,U\sum_{{\bf k'},i\nu'} \partial_{\Lambda} G^{\Lambda}({\bf k'},i\nu')  e^{\pm i \nu 0^+}   \label{eq:sigma_G}\\
    \dot{\Sigma}_{GGG}({\bf k},i\nu)=& 
    - U^2 \sum_{{\bf k' q}}\sum_{i\nu' i\omega}
    \partial_{\Lambda}\Big[G^{\Lambda}({\bf k'},i\nu')
    G^{\Lambda}({\bf k}+{\bf q},i\nu+i\omega) 
    G^{\Lambda}({\bf k'}+{\bf q},i\nu'+i\omega)\Big]  \label{eq:sigma_GGG}\\
    \dot{\Sigma}_{pp}({\bf k},i\nu) =&-\sum_{{\bf k'}i\nu'} \sum_{m}f^*_m({\bf k})  f_{0}({\bf k}) 4\pi^2 U  \nonumber \\
    &\partial_{\Lambda} \Big[ \sum_{i\nu''} \sum_{n}
    \Phi^{\Lambda}_{pp}({\bf k'}+{\bf k},m,n,i\nu'+i\nu,i\nu,i\nu'')
    \Pi^{\Lambda}_{pp}({\bf k'}+{\bf k},n,0,i\nu'+i\nu,i\nu'') 
    G^{\Lambda}({\bf k'},i\nu') \Big] \;, \label{eq:sigma_pp} \\
    \dot{\Sigma}_{ph}({\bf k},i\nu) =& -\sum_{{\bf k'}i\nu'} \sum_{m}f^*_m({\bf k}) f_{0}({\bf k}) 4\pi^2 U  \nonumber \\
    &\partial_{\Lambda} \Big[ \sum_{i\nu''} \sum_{n}
    \Phi^{\Lambda}_{ph}({\bf k'}-{\bf k},m,n,i\nu'-i\nu,i\nu,i\nu'')
    \Pi^{\Lambda}_{ph}({\bf k'}-{\bf k},n,0,i\nu'-i\nu,i\nu'') 
    G^{\Lambda}({\bf k'},i\nu') \Big]  \label{eq:sigma_ph}\\
    \dot{\Sigma}_{\overline{ph}}({\bf k},i\nu) =& -\sum_{{\bf k'}i\nu'} \sum_{m}f^*_m({\bf k}) f_{0}({\bf k}) 4\pi^2 U \nonumber \\
    &\partial_{\Lambda} \Big[ \sum_{i\nu''} \sum_{n}
    \Phi^{\Lambda}_{\overline{ph}}({\bf k'}-{\bf k},m,n,i\nu'-i\nu,i\nu,i\nu'')
    \Pi^{\Lambda}_{ph}({\bf k'}-{\bf k},n,0,i\nu'-i\nu,i\nu'') 
    G^{\Lambda}({\bf k'},i\nu') \Big] \label{eq:sigma_xph} \;,
\end{align}
\end{widetext}
with
\begin{subequations}
\begin{align}
    \Pi_{ph}({\bf q},n,m,i\omega,i\nu)&= \int d{\bf p} f^*_n({\bf p}) f_{m}({\bf p}) G({\bf p},i\nu) \nonumber \\
    & \hspace{0.7cm} G({\bf q}+ {\bf p},i\omega+i\nu) \\
    \Pi_{pp}({\bf q},n,m,i\omega,i\nu)&= \int d{\bf p} f^*_n({\bf p}) f_{m}({\bf p}) G({\bf p},i\nu) \nonumber \\
    & \hspace{0.7cm} G({\bf q} - {\bf p},i\omega - i\nu) \;.
\end{align}
\end{subequations}
If not otherwise specified, we use the single-scale propagator  $\partial_{\Lambda}G^{\Lambda}\approx S^{\Lambda}=\partial_{\Lambda}G^{\Lambda}|_{\Sigma=\textrm{const}}$.
Note that in Eqs.~\eqref{eq:sigma_GGG}-\eqref{eq:sigma_xph}, the derivative with respect to $\Lambda$ on the right hand sides yields three contributions each. 
These depend on the ($1\ell$) flow of the two-particle vertex \cite{TagliaviniHille2019}, that supplements the above Eqs.~\eqref{eq:sigma_G}-\eqref{eq:sigma_xph}.
In the flow equations for the vertex, the self-energy $\Sigma (k)$ is inserted into the full Green's functions on the right hand sides without further expansions around the Fermi surface or in small frequencies.

We observe that in principle one could also use directly the SDE \eqref{eq:SDE} replacing its scale derivative in Eqs.~\eqref{eq:sigma_G}-\eqref{eq:sigma_xph}.
However, we preferred a formulation which remains as close as possible to the original fRG idea involving a differential flow equation for the self-energy.  Besides the more straightforward numerical implementation of the derivative in an adaptive differential equation solver, we believe that the close relation to the multiloop fRG flow by which the proposed SDE scheme is inspired makes it more intuitive. In fact, including the Katanin correction or any other higher loop order as discussed in Sec.~\ref{ssec:mfRG}, requires anyways the computation of the self-energy derivative with Eqs.~\eqref{eq:sigma_G}-\eqref{eq:sigma_xph}. 

For the two-particle vertex \footnote{For the convention on the flow 
equations, see Ref.~\onlinecite{HilleThesis}.}, we consider only a single local $s$-wave form factor and a small number of frequencies, verifying that an additional $d$-wave form factor as well as more frequencies do not qualitatively affect the results in the considered parameter regime. Specifically, we use $8$ fermionic and $17$ bosonic frequencies in the low frequency object depending on all three frequencies. The same numbers are used in the high-frequency asymptotics of a single fermionic frequency for the remaining bosonic and fermionic frequencies. The asymptotic of both fermionic frequencies is described by $513$ bosonic frequencies. 
Concerning the transfer momentum parametrization, in addition to $16 \times 16$ momentum patches distributed on an equally spaced grid in the Brillouin zone, we take into account a finer $5 \times 5$ grid around the AF peak at $\bf{q}=(\pi,\pi)$, see also Fig.~\ref{fig:Gapopening_vs_KD} for a more detailed convergence analysis.

We finally note that we use here the so-called interaction ``cutoff" with $G^{0,\Lambda}=\Lambda G_0$ which gradually turns on the bare onsite interaction~$U$~\cite{Honerkamp2004}. 
This allows one to translate the flowing self-energy and two-particle vertex at scale $\Lambda$ to the solution (at the end of the flow) of a rescaled bare vertex $\Lambda^2U$, see Appendix~\ref{app:scale_intflow} for the proof.
This cutoff choice is motivated by the possibility to trace the onset of the gap opening along the flow and identify the pseudo-critical interaction in correspondence of the vertex divergence. The absolute values depend on the applied flow scheme and for this reason a quantitative comparison to other cutoffs appears difficult. At the same time, the analytical arguments as well as our findings at higher loop order presented below (see also Ref.~\onlinecite{TagliaviniHille2019} for a more detailed discussion on the cutoff independence of the multiloop fRG) support the robustness of the qualitative differences between the conventional and the SDE flow. Therefore, it should be observed independently of the flow scheme.

\section{Results}

\subsection{Self-energy flow versus Schwinger-Dyson equation}
\label{sec:1l_vs_SDE}

We present here results for the self-energy together with an analysis of the differences between the fRG calculation using the $1\ell$ flow Eq.~\eqref{eq:Sigma_1l} with respect to the derivative of the SDE in Eq. \eqref{eq:sigmasplit}. In particular, we show how in the latter the long-range AF fluctuations lead to a gap opening, see also Appendix~\ref{app:xph}.

\begin{figure}[b]
    \centering
    \includegraphics[width=\cw]{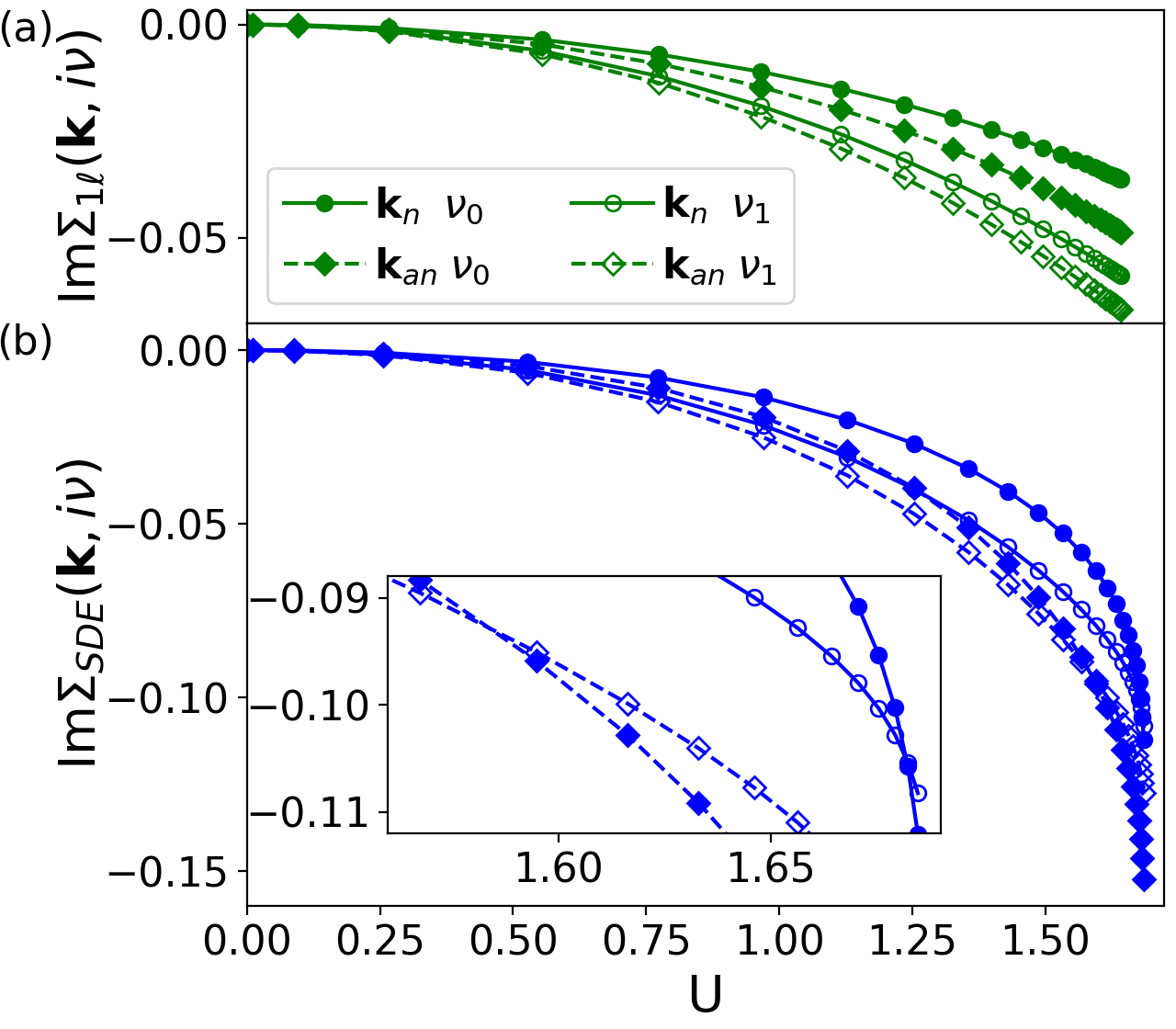}
    \caption{Self-energy as a function of the flowing interaction $U$ for $1/T=10$. Comparison of the nodal (circles, solid lines) and antinodal points (diamonds, dashed lines) for the first and second Matsubara frequencies (closed and open symbols respectively) (a) in the conventional $1\ell$ scheme and (b) with the derivative of the SDE for the self-energy. The crossings in the latter (inset) indicate the gap opening, occurring first at the antinodal and then at the nodal point.}
    \label{fig:Sigma_vs_U}
\end{figure}

Let us first look at the flow of the self-energy displayed in Fig.~\ref{fig:Sigma_vs_U}, for an inverse temperature of $1/T=10$. The panel (a) corresponds to the fRG calculation with the $1\ell$ flow of the self-energy and (b) to the scheme with the derivative of the SDE. We follow the flow at the nodal ${\bf k}_n=(\pi/2,\pi/2)$ (circles) and antinodal ${\bf k}_{an}=(\pi,0)$ (diamonds, solid lines) momentum point, both for the first (closed symbols) and second (open symbols, dashed lines) Matsubara frequencies. 
The imaginary part of the self-energy is shown as a function of the flowing interaction $U$.
The end point on the right defines the so-called pseudo-critical interaction at which the maximal component in one of the channels in the vertex exceeds $10^{3}$. Due to the self-energy feedback into the flow of the vertex, and vice versa, it depends on the flow scheme applied to the self-energy.
In the SDE scheme this divergence sets in at $U=1.69$, which is larger than in $1\ell$ where the flow diverges at $U=1.64$. 

This is consistent as the self-energy is larger in the SDE scheme and therefore its screening of the dominating $\overline{ph}$-excitations in the 2D Hubbard model at half filling
is stronger (see also Ref.~\onlinecite{TagliaviniHille2019}).
In the inset of (b) a zoom on the crossings of the imaginary parts of the self-energy at the first and second Matsubara frequency is shown, which at sufficiently low temperatures is associated with a smooth, non-critical transition between a Fermi-liquid and an insulating behavior \cite{Simkovic2020}. 
In particular, this transition occurs first at the antinodal momentum point and only for large values of the bare interaction at the nodal point. 
The region in between identifies the pseudogap regime which we will discuss in the following.

\begin{figure}[t]
    \centering
    \includegraphics[width=\cw]{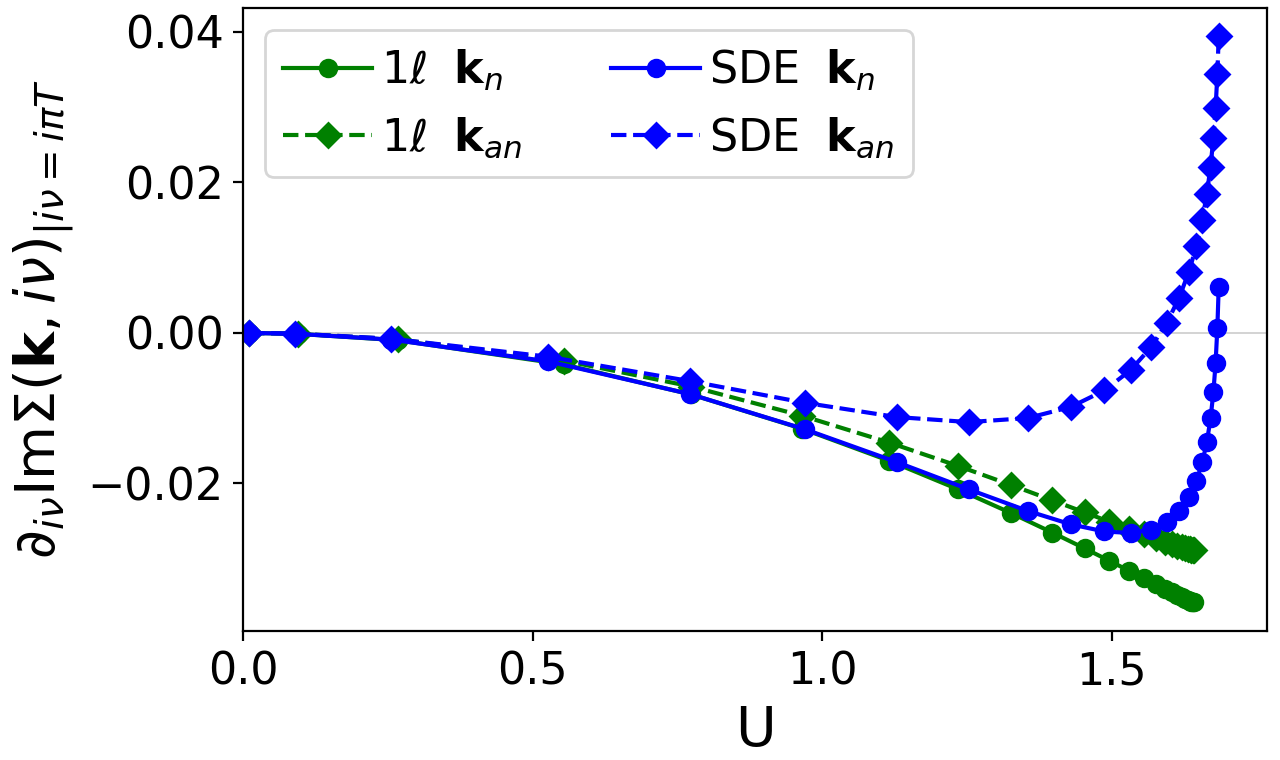}
    \caption{$\partial_{i\nu}\operatorname{Im}\Sigma({\bf k},i\nu)$ evaluated at $i\nu=i\pi T$ as a function of the flowing interaction $U$ for $1/T=10$. The conventional $1\ell$ self-energy flow (green) is compared to the one of the derivative of the SDE (blue). The latter crosses zero for both momentum points indicating the opening of a gap, while the conventional flow exhibits a monotonic behavior.}
    \label{fig:Zfactor_flow_vs_SDE}
\end{figure}

The presence of quasiparticles in a Fermi-liquid is equivalent to a nonzero quasiparticle weight \cite{Mahan2000} 
\begin{align} \label{eq:QP_weight}
    Z({\bf k})=\Big( 1-\frac{\partial \operatorname{Re}\Sigma(\nu,{\bf k})}{\partial \nu}\Big|_{\nu \rightarrow 0} \Big)^{-1} \;,
\end{align}
where $\nu$ is a real frequency. 
In a Fermi liquid, we have a nonzero $Z({\bf k}) < 1$. Non-Fermi-liquid behavior can be signaled by deviations from this, e.g. $Z({\bf k}) \to 0$ or $Z ({\bf k})> 1$. $Z ({\bf k})\to 0$ amounts to an infinitely steep slope of 
$\operatorname{Re}\Sigma(\nu,{\bf k})$ at $\nu = 0 $ and thus a vanishing of the quasi-particle weight \emph{without} any dip or pseudogap feature in the spectral function, while $Z ({\bf k})> 1$ corresponds to a positive slope of $\operatorname{Re}\Sigma(\nu,{\bf k})$ at $\nu = 0$ and implies the emergence of a double-peak structure of the spectral function when this slope is large enough. In the low temperature limit, $\partial_{\nu} \operatorname{Re}\Sigma(\nu,{\bf k})|_{\nu \rightarrow 0}$ can be translated to Matsubara frequencies. Then, the gap opening can be observed directly in the imaginary part of the self-energy. For a Fermi liquid, the imaginary part goes linearly through zero imaginary frequency. 
In the gapped regime, $\operatorname{Im}\Sigma$ bends towards positive large values approaching the zero (Matsubara) frequency limit from below and towards negative large values from above while in the Fermi liquid regime the bending is always towards small values. 
The onset of a pseudo-gap can hence be detected by the crossing of the imaginary parts of the self-energy at the first and second Matsubara frequency (see also Refs.~\cite{Schaefer2015,Schaefer2016,Simkovic2020} and~\cite{Schaefer2020}).
Therefore, we study the difference between the latter via
\begin{align} \label{eq:factor}
    \partial_{i\nu}\operatorname{Im}\Sigma({\bf k},i\nu)|_{i\nu=i\pi T}=\frac{\operatorname{Im}\Sigma({\bf k},i3\pi T)-\operatorname{Im}\Sigma({\bf k},i\pi T)}{2\pi T} \;,
\end{align}
with $\partial_{i\nu}\operatorname{Im}\Sigma({\bf k},i\nu)|_{i\nu=i\pi T}\le0$ corresponding to the Fermi-liquid-like regime and  $\partial_{i\nu}\operatorname{Im}\Sigma({\bf k},i\nu)|_{i\nu=i\pi T}>0$ to a pseudogap at momentum $\bf{k}$. 

\begin{figure}[t]
    \centering
    \includegraphics[width=\cw]{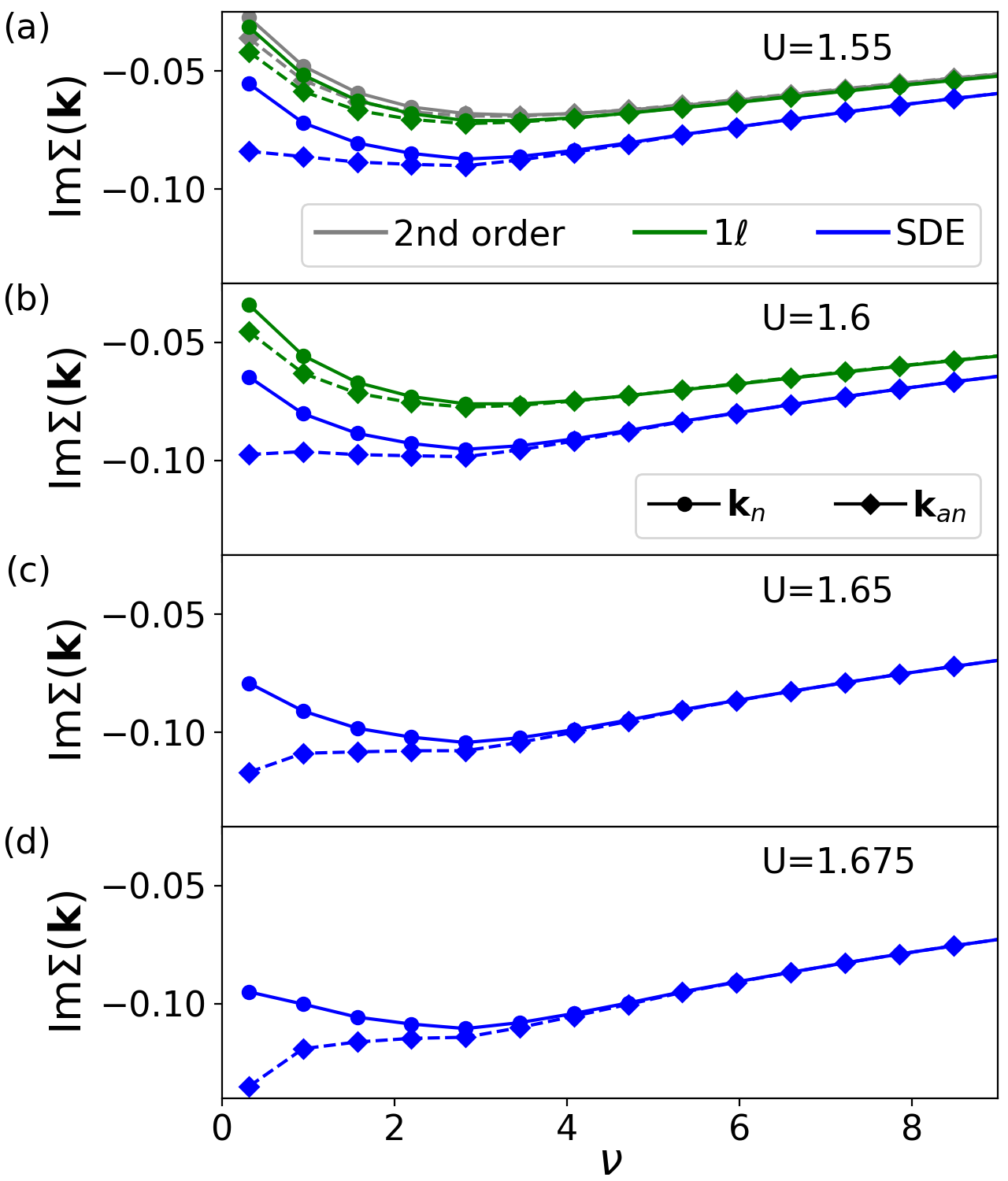}
    \caption{Self-energy as a function of the Matsubara frequency $i\nu_n$ for $1/T=10$ and (a) $U=1.55$, (b) $U=1.6$, (c) $U=1.65$ and (d) $U=1.675$ just below and above the gap opening at the nodal (circles) and antinodal (diamonds) points. The results obtained in the conventional $1\ell$ flow are shown in green, the ones from the derivative of the SDE in blue. Note that for $U=1.65$ the $1\ell$ flow is already diverged. For comparison also the second-order perturbation theory for $U=1.55$ is shown in gray. 
    }
    \label{fig:Sig_iw_T0p1}
\end{figure}

We report $\partial_{i\nu}\operatorname{Im}\Sigma({\bf k},i\nu)|_{i\nu=i\pi T}$ in Fig.~\ref{fig:Zfactor_flow_vs_SDE}, again for both flow schemes.  
The zeros corresponding to $\partial_{i\nu}\operatorname{Im}\Sigma({\bf k},i\nu)|_{i\nu=i\pi T}=0$ in the SDE scheme (corresponding to the crossings observed in Fig.~\ref{fig:Sigma_vs_U}) can be directly read off. The gap opening at the antinodal momentum point sets in first, followed by the one at the nodal point close to the vertex divergence.
Note that the $1\ell$ flow does not exhibit any zeros. Using the derivative of the SDE replacing the conventional $1\ell$ flow of the self-energy is therefore crucial for the description of spectral properties, i.e. for seeing the gap open up.

A more conventional representation of the self-energy as a function of the Matsubara frequency is provided in Fig.~\ref{fig:Sig_iw_T0p1}, for bare interactions close to the gap opening. Below the gap opening (for $U=1.55$), we observe a Fermi-liquid behavior with a pronounced upturn towards zero at low frequencies in both the SDE and the $1\ell$ flow.
At the antinodal momentum point, the self-energy resulting from the SDE flow already exhibits a tendency to a non Fermi-liquid behavior.
Once the gap is opened, this behavior turns first into a slight (for $U=1.6$) and then to a more pronounced (for $U=1.65$) downturn at the first Matsubara frequency, while the $1\ell$ results remain Fermi-liquid like for all values of $U$ (for comparison also the second-order perturbation theory is shown).
At the nodal point, we still find a Fermi-liquid behavior in agreement with a picture in which the pseudogap opens at the anti-nodal points first.
For the observation of the gap opening at the nodal point, we would have to tune the bare interaction very close to the pseudo-critical interaction.

\begin{figure}[t]
    \centering
    \includegraphics[width=\cw]{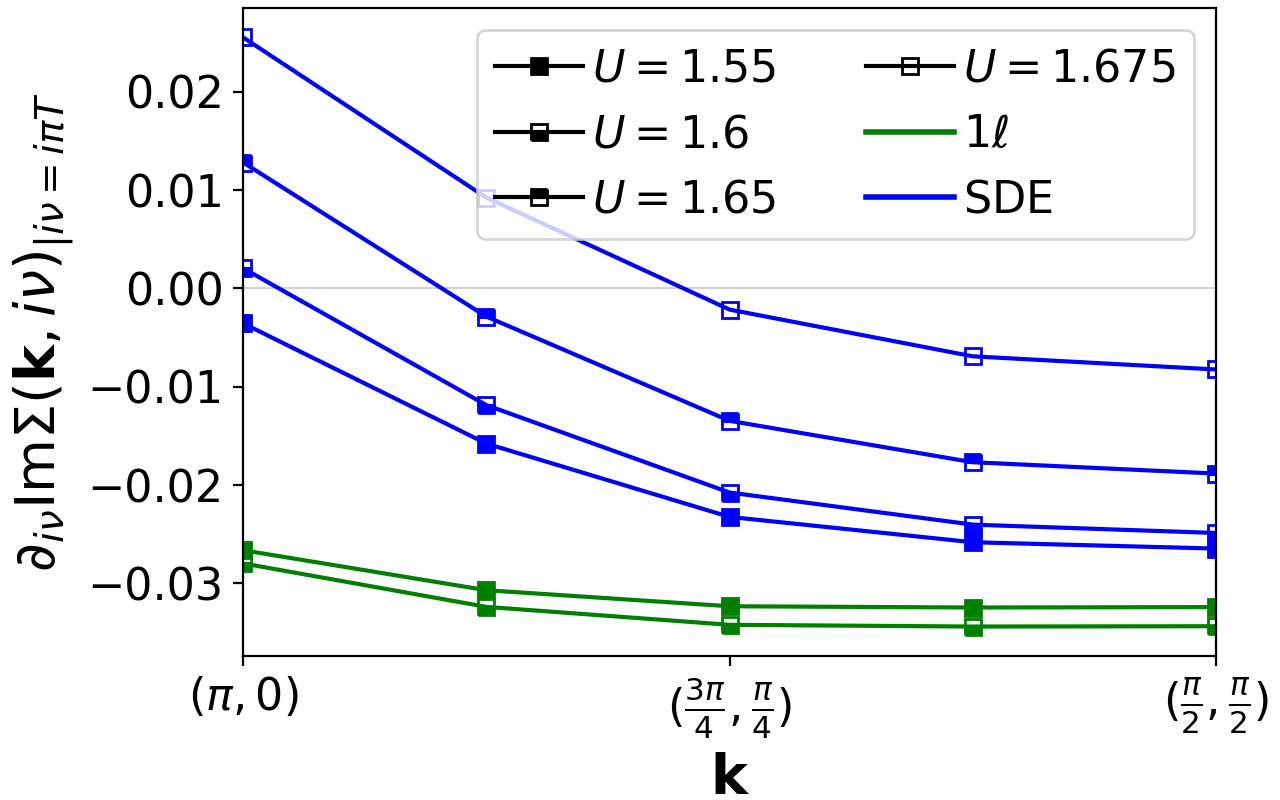}
    \caption{$\partial_{i\nu}\operatorname{Im}\Sigma({\bf k},i\nu)$ evaluated at $i\nu=i\pi T$ as a function of the momentum on the Fermi surface.}
    \label{fig:Zfactor_vs_k}
\end{figure}

The momentum anisotropy can also be studied in the evolution of $\partial_{i\nu}\operatorname{Im}\Sigma({\bf k},i\nu)|_{i\nu=i\pi T}$ along the Fermi surface, shown in Fig.~\ref{fig:Zfactor_vs_k} for the same parameters. We note that the spread with momentum is significantly larger in the SDE flow where the monotonic decrease from the antinodal to the nodal momentum point crosses zero and opens the gap with increasing bare interaction first for $U=1.6$ in proximity of the antinodal point.
The $1\ell$ fRG flow instead leads to Fermi-liquid behavior on the whole Fermi surface. 

\begin{figure}[b]
    \centering
    \includegraphics[width=\cw]{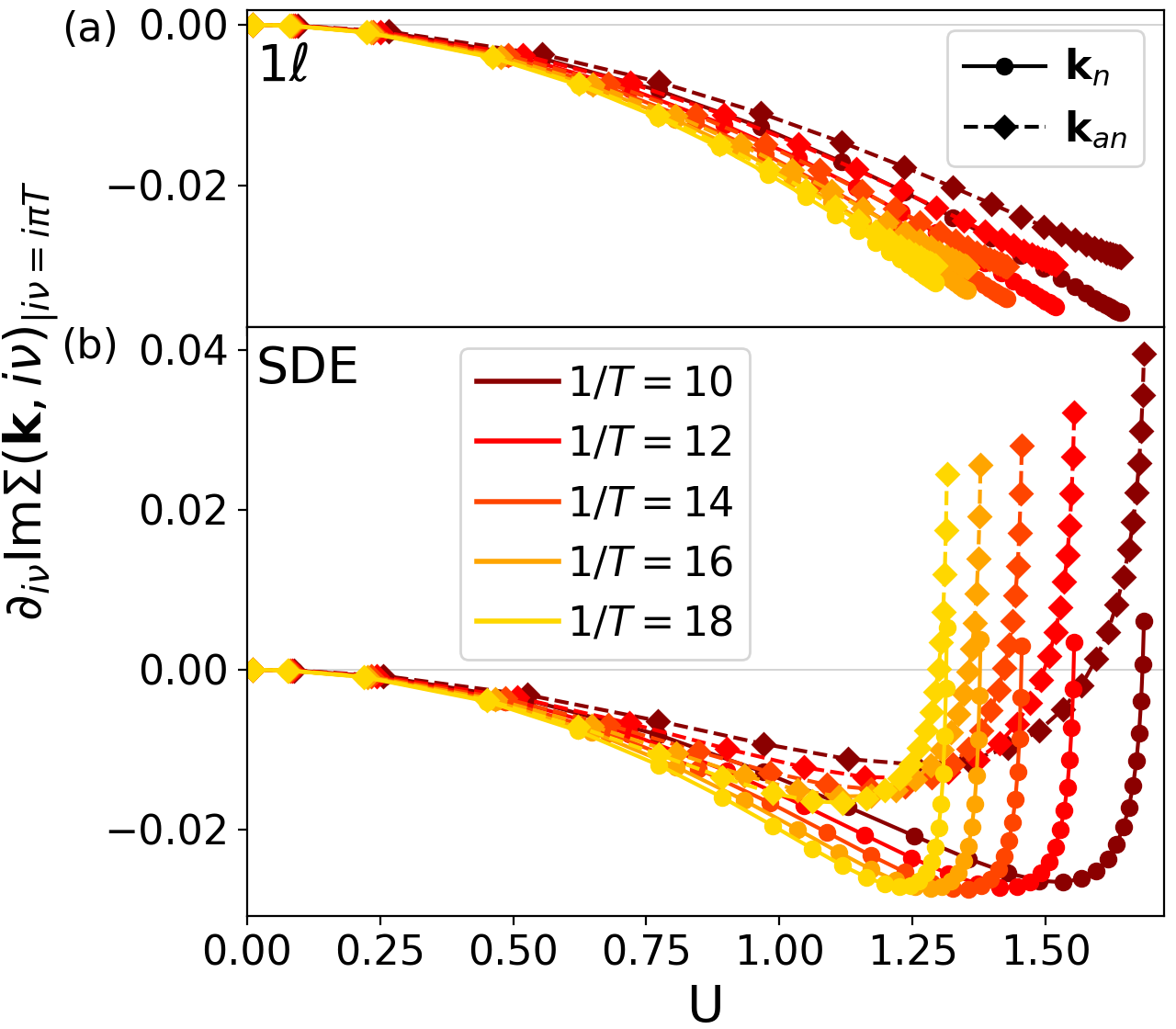}
    \caption{$\partial_{i\nu}\operatorname{Im}\Sigma({\bf k},i\nu)$ evaluated at $i\nu=i\pi T$ as a function of the flowing interaction $U$ for different temperatures. Using the (a) conventional $1\ell$ self-energy flow, no gap opening occurs for any temperature, while (b) the derivative of the SDE yields a gap opening which for decreasing $T$ sets in at lower values of the effective $U$. 
    }
    \label{fig:Zfactor_temperatures}
\end{figure}

\begin{figure}[t]
    \centering
    \includegraphics[width=\cw]{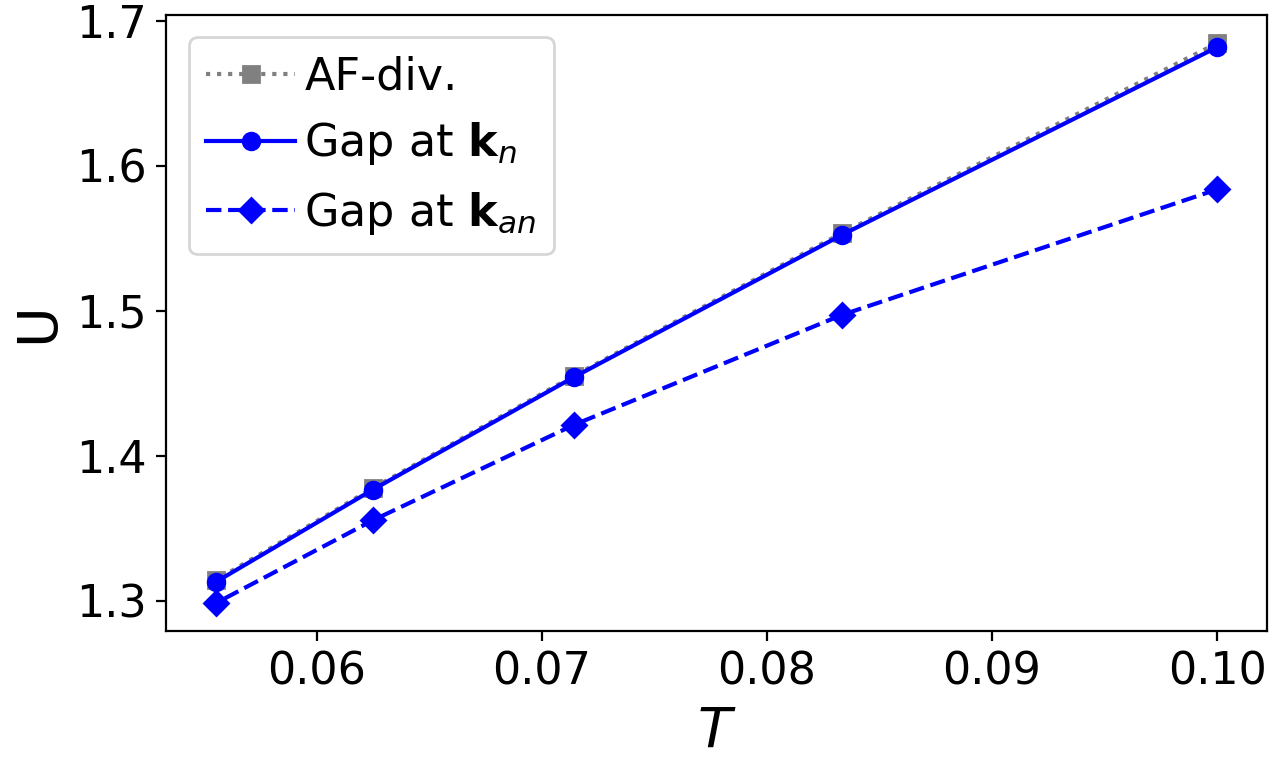}
    \caption{Flowing interaction $U$ at which the gap opens as a function of the temperature. The gap opening is shown for the antinodal point (blue diamonds) and the nodal one (blue dots), which occurs in the proximity of the AF divergence (gray squares).}
    \label{fig:Gapopening_vs_T}
\end{figure}

We now investigate the temperature dependence of the gap opening, starting from the considered $1/T=10$ in Figs. \ref{fig:Sigma_vs_U}-\ref{fig:Zfactor_vs_k} down to $1/T=18$, see Fig.~\ref{fig:Zfactor_temperatures} for the respective results for $\partial_{i\nu}\operatorname{Im}\Sigma({\bf k},i\nu)|_{i\nu=i\pi T}$. The detected behavior is qualitatively the same: no gap opening occurs for any temperature in the $1\ell$ flow, whereas the derivative of the SDE yields a gap which sets in first at the antinodal and subsequently at the nodal point.
For decreasing $T$, the pseudo-critical interaction and with it the gap opening is shifted to lower values, reducing at the same time the distance between the zeros of $\partial_{i\nu}\operatorname{Im}\Sigma({\bf k},i\nu)|_{i\nu=i\pi T}$ at the antinodal and nodal point and the pseudo-critical scale. This can be seen in Fig.~\ref{fig:Gapopening_vs_T}, where we show the temperature dependence of the interaction at which the gaps open and of the pseudo-critical interaction (gray squares) at which the $\overline{ph}$-channel exceeds the critical value and the flow is stopped.  
We find that the antinodal gap opening is clearly distinct from the pseudo-critical interaction even for the lowest temperature considered here. In contrast, the pseudo-critical interaction almost coincides with the gap opening at the nodal point that sets in only a little below. 

\begin{figure}[t]
    \centering
    \includegraphics[width=\cw]{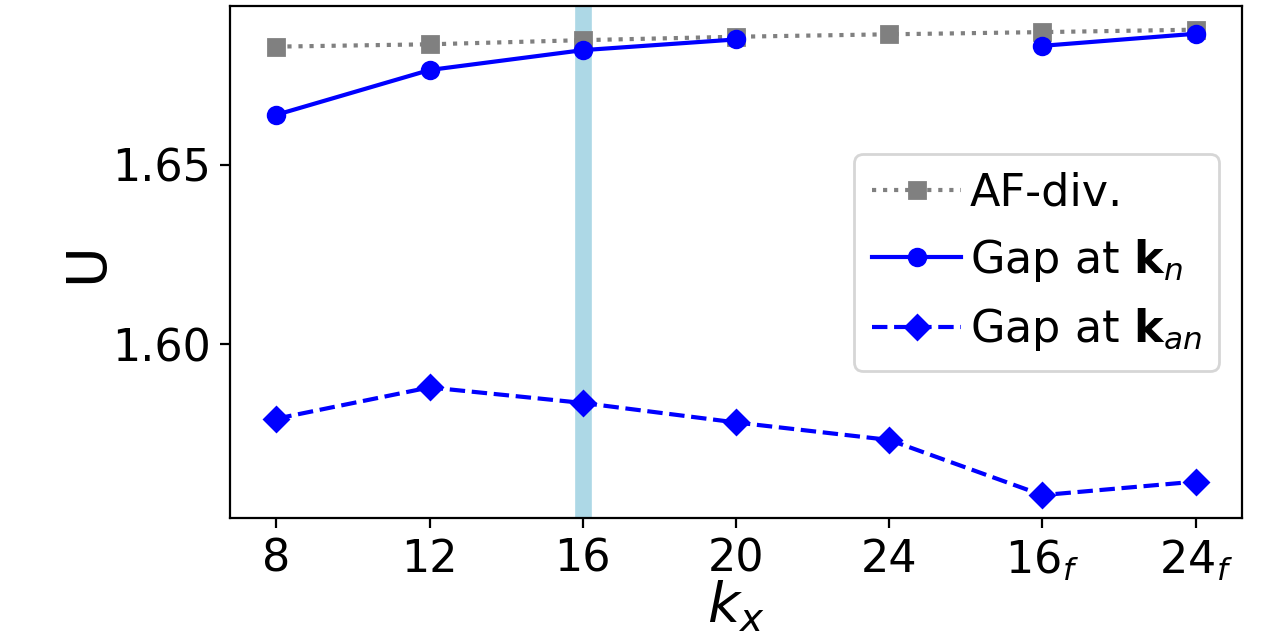}
    \caption{Flowing interaction $U$ at which the gap opens as a function of the number of momenta $k_x$ (with $k_y=k_x$ and subscript $f$ for enlarged fine patching region) accounted for in the two-particle vertex, for $1/T=10$.  
    The gap at the antinodal point (blue diamonds) opens always before the AF divergence (grey squares) sets in, while the gap at the nodal point (blue dots) vanishes with increasing resolution of the Brillouin zone, see text for details. The light blue line corresponds to the parametrization of all other computations.}
    \label{fig:Gapopening_vs_KD}
\end{figure}

At the same time, the presented fRG data is sensitive to the employed parametrization of the two-particle vertex. In particular, a convergence in frequencies (with minor impact, see discussion in Sec.~\ref{ssec:fRG}) and momenta becomes numerically challenging for lower values of $T$ and is not fully reached yet.
The latter is illustrated in Fig.~\ref{fig:Gapopening_vs_KD}. 
Upon increasing the number of patches, the gap opening at the nodal point is shifted closer towards the pseudo-critical interaction until it eventually merges with the latter around 20 $k_x$ momentum points. With a further refinement around $\bf{q}=(\pi,\pi)$ including $15\times15$ instead of the former $5\times5$ patches and covering a larger momentum area (indicated by the subscript $f$ on the right side of Fig.~\ref{fig:Gapopening_vs_KD}), the gap opening can be resolved again as a precursor of the pseudo-criticality. In contrast, the gap opening at the antinodal point does not approach the pseudo-critical interaction. 
The disappearance of the nodal gap as a precursor is due to the decreasing region covered by the fine patching and the consequently reduced resolution of the AF peak for larger values of $k_x$, i.e. the region becomes smaller than the peak width. When turning to a larger fine-patching region  (indicated by the subscript $f$), the peak width is again covered, explaining the nonmonotonic behavior.

Hence, within the present analysis, the question whether in the low-temperature regime the gap opening at the nodal point is stable as a distinct feature from the occurrence of long-range correlations cannot be answered conclusively. 
Nevertheless, the data shown here are consistent with the physical picture emerging from various other state-of-the-art many-body techniques for the weak- or moderate-coupling regime of the 2D Hubbard model at half filling, as, e.g., in \cite{Huscroft2001,Senechal2004,Gull2010}, or currently collected in \cite{Schaefer2020}.
The AF correlations (associated to an exponentially increasing correlation length) increase at lower temperatures and eventually diverge at $T = 0$ when the ground-state characterized by an AF long-range order is reached. In this low-temperature regime, long-wavelength AF fluctuations lead to an enhanced quasiparticle scattering rate and to the formation of a pseudogap in the single-particle spectrum (which evolves into a sharp gap in the Slater-like insulator at $T = 0$). These fluctuations gradually destroy the coherent quasiparticles of the Fermi liquid. 
The crossover temperature corresponding to the pseudogap opening is not uniform along the Fermi surface: it is higher at the antinodal and lower at the nodal points. Eventually, all states of the Fermi surface are suppressed by AF fluctuations, resulting in a full gap. 

\subsection{Towards full multiloop fRG}
\label{ssec:mfRG}

\begin{figure}[ht]
    \centering
    \includegraphics[width=\cw]{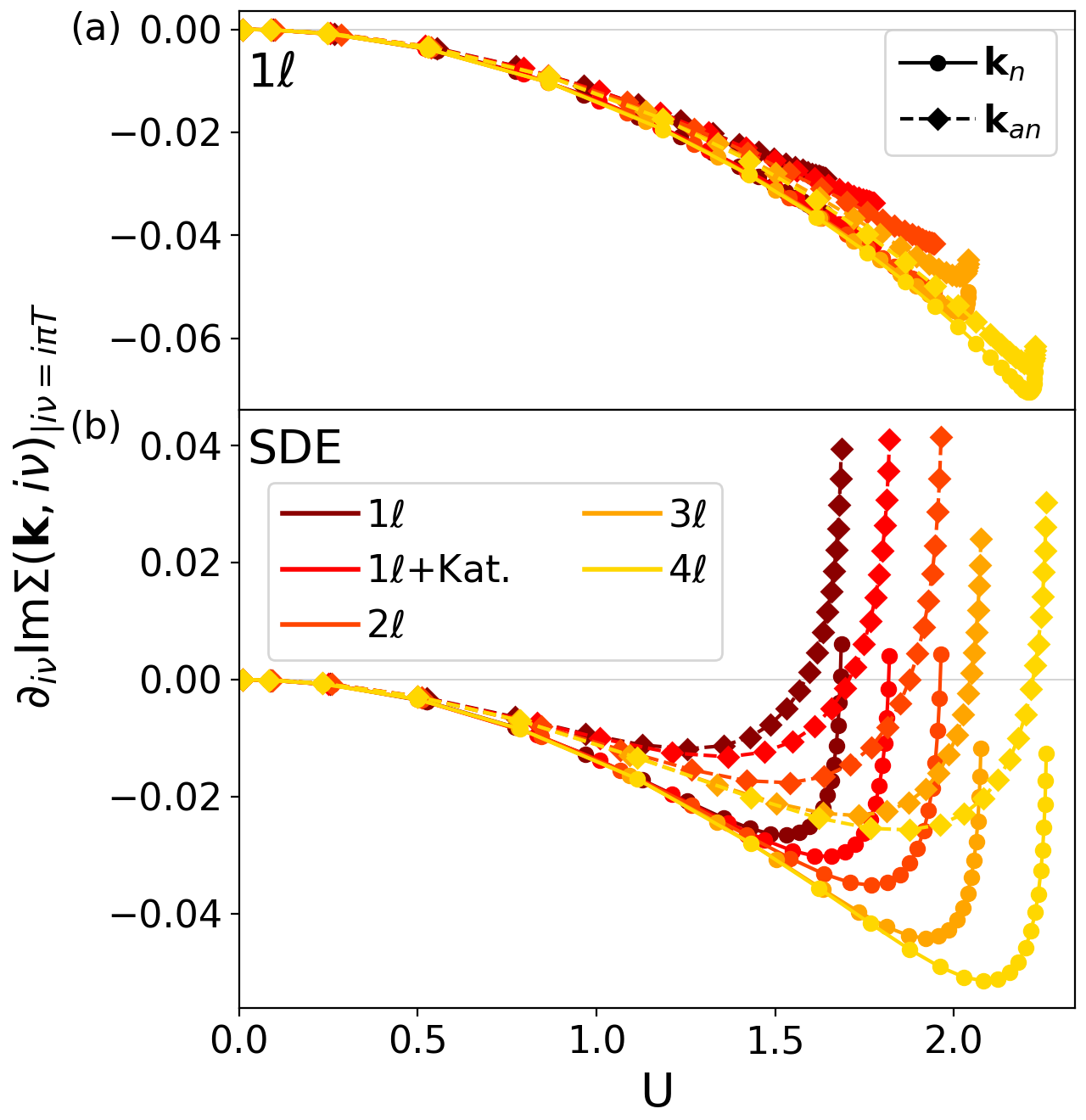}
    \caption{$\partial_{i\nu}\operatorname{Im}\Sigma({\bf k},i\nu)$ evaluated at $i\nu=i\pi T$ as a function of the flowing interaction $U$, for $1/T=10$ and different loop orders $\ell$, in both (a) the conventional fRG and (b) the SDE-approach for the self-energy flow which takes into account the form-factor projections in the different channels. In the conventional fRG a tendency towards gap opening is observed only very close to the pseudo-critical temperature without actually leading to a crossing of the self-energy in the first Matsubara frequencies. In the SDE-approach, the gap opening at the antinodal point is observed also at higher loop order, while the gap at the nodal point vanishes with increasing $\ell$. For a direct comparison of the gap opening scales and the onset of the AF divergence we refer to Fig.~\ref{fig:Gapopening_vs_loop} including higher loop orders. 
    }
    \label{fig:Zfactor_loops}
\end{figure}

\begin{figure}[ht]
    \centering
    \includegraphics[width=\cw]{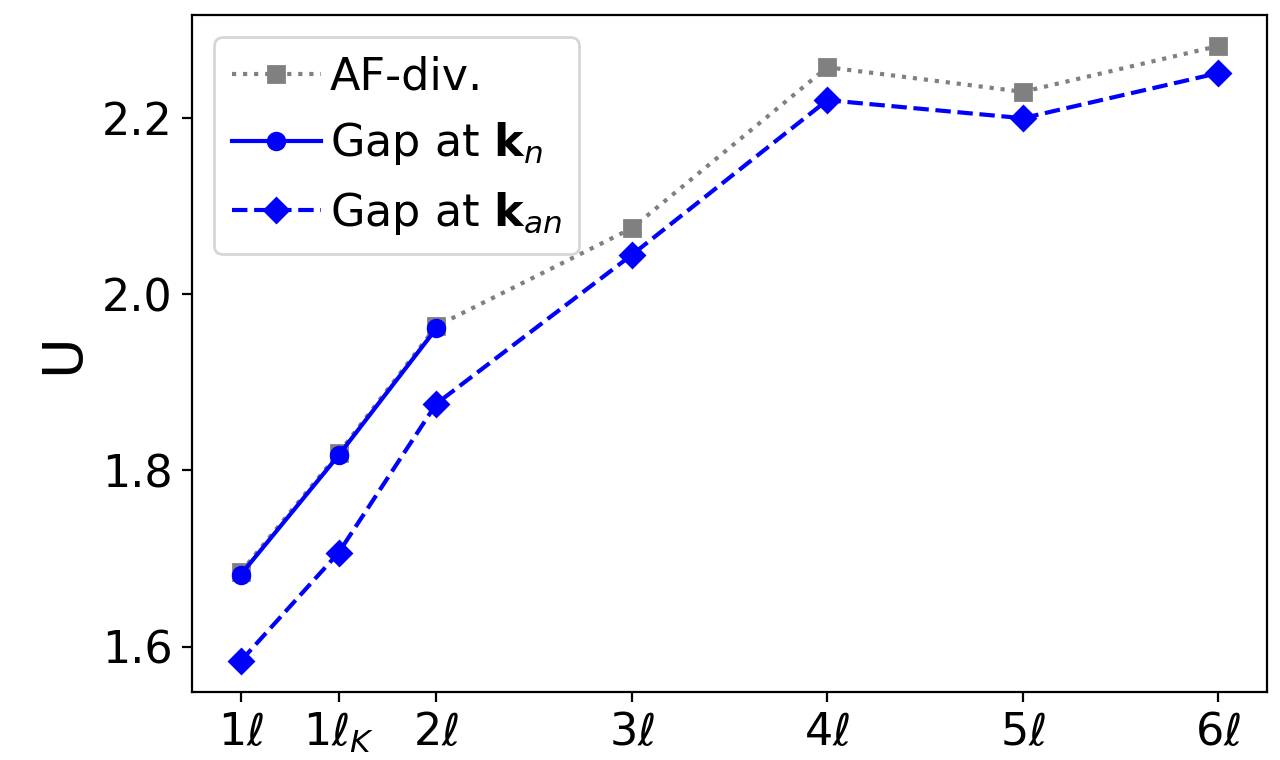}
    \caption{Flowing interaction $U$ at which the gap opens as a function of the loop order $\ell$, for $1/T=10$. The gap opening at the antinodal point (blue diamonds) occurs always before the AF divergence (grey squares), while the gap opening at the nodal point (blue dots) disappears into the pseudo-ordered phase at higher loop order. The oscillatory behavior is characteristic for the loop convergence  \cite{TagliaviniHille2019}.}
    \label{fig:Gapopening_vs_loop}
\end{figure}

The $1\ell$ description truncates the infinite hierarchy of flow equations \cite{Metzner2012} after the two-particle vertex. In the multiloop extension \cite{Kugler2018a,Kugler2018b}, a part of the higher order vertex contributions is taken into account effectively by including higher loop contributions to the one- and two-particle vertex flows. The inclusion of all loop orders yields the parquet approximation enhancing the pseudo-critical interactions and lowering the pseudo-critical temperatures \cite{TagliaviniHille2019} with respect to the $1\ell$ truncation. According to the Mermin-Wagner theorem \cite{Mermin1966}, which is fulfilled by the parquet approximation, the instability eventually disappears (this regime is however extremely difficult to reach numerically). 
We note that at infinite loop order, the results no longer depend on the employed flow scheme. We hence expect the trend observed for the interaction flow to be qualitatively reproduced also by other flow schemes at finite loop order. 

For the precise form of the multiloop equations, we refer to Refs.~\cite{Kugler2018b} and~\cite{TagliaviniHille2019}. In the SDE-scheme for the self-energy, we replaced both the $1\ell$ equation and multiloop corrections of the self-energy by Eq.~\eqref{eq:SDE}.
Note that for the Katanin replacement and in any higher loop order the differentiated propagator becomes $\partial_{\Lambda}G^{\Lambda}=S^{\Lambda}+G^{\Lambda}\dot{\Sigma}^{\Lambda}G^{\Lambda}$. For the second part of $\partial_{\Lambda}G^{\Lambda}$, the self-energy change has to be known before the calculation of the vertex flow and, in the SDE-scheme, inside the self-energy flow itself. Here, we replace it with the $1\ell$ flow in Eq.~\eqref{eq:Sigma_1l} independently of the self-energy scheme used. For a full feedback of the self-energy change according to Eq.~\eqref{eq:sigmasplit} to the Katanin replacement, further iterations within the same $\Lambda$-step should be performed. As the correction is only of quantitative nature \cite{Hille2020}, we neglect here these iterations.

In Fig.~\ref{fig:Zfactor_loops} we present the results for $\partial_{i\nu}\operatorname{Im}\Sigma({\bf k},i\nu)$ evaluated at $i\nu=i\pi T$ as a function of the flowing interaction $U$, for $1/T=10$ and different loop orders $\ell$. While in the conventional fRG no gap opens at any loop order, in the SDE-approach, the gap opening at the antinodal point is observed also at higher loop order, before the flow has to be stopped and pseudo-criticality is reached. 
In Appendix~\ref{app:scale_intflow}, we show that also for the multiloop flow equations, the scale at which the interaction flow diverges can be translated into a pseudo-critical interaction.

A direct comparison of the gap opening interaction and the onset of the AF divergence is shown in Fig.~\ref{fig:Gapopening_vs_loop}, as a  function of the loop  order (for  $1/T=10$). 
A trend towards higher pseudo-critical interactions can be observed already at the first loop orders, while the oscillatory behavior is characteristic for the loop convergence \cite{TagliaviniHille2019}.
The larger pseudo-critical interactions could in principle leave more space for the pseudogap to develop. However, the gap at the nodal point vanishes at higher loop order, while at the antinodal point sets in at a rather small but constant distance from the pseudo-critical line. 
Therefore, the gap opening tendency is not loop-order dependent and the only remaining difference between the $1\ell$ and SDE scheme is the form-factor truncation.
Further, this indicates indirectly that the gap opening is driven by strong AF fluctuations, see also Appendix \ref{app:xph}.

\begin{figure}[b]
    \centering
    \includegraphics[scale=0.4]{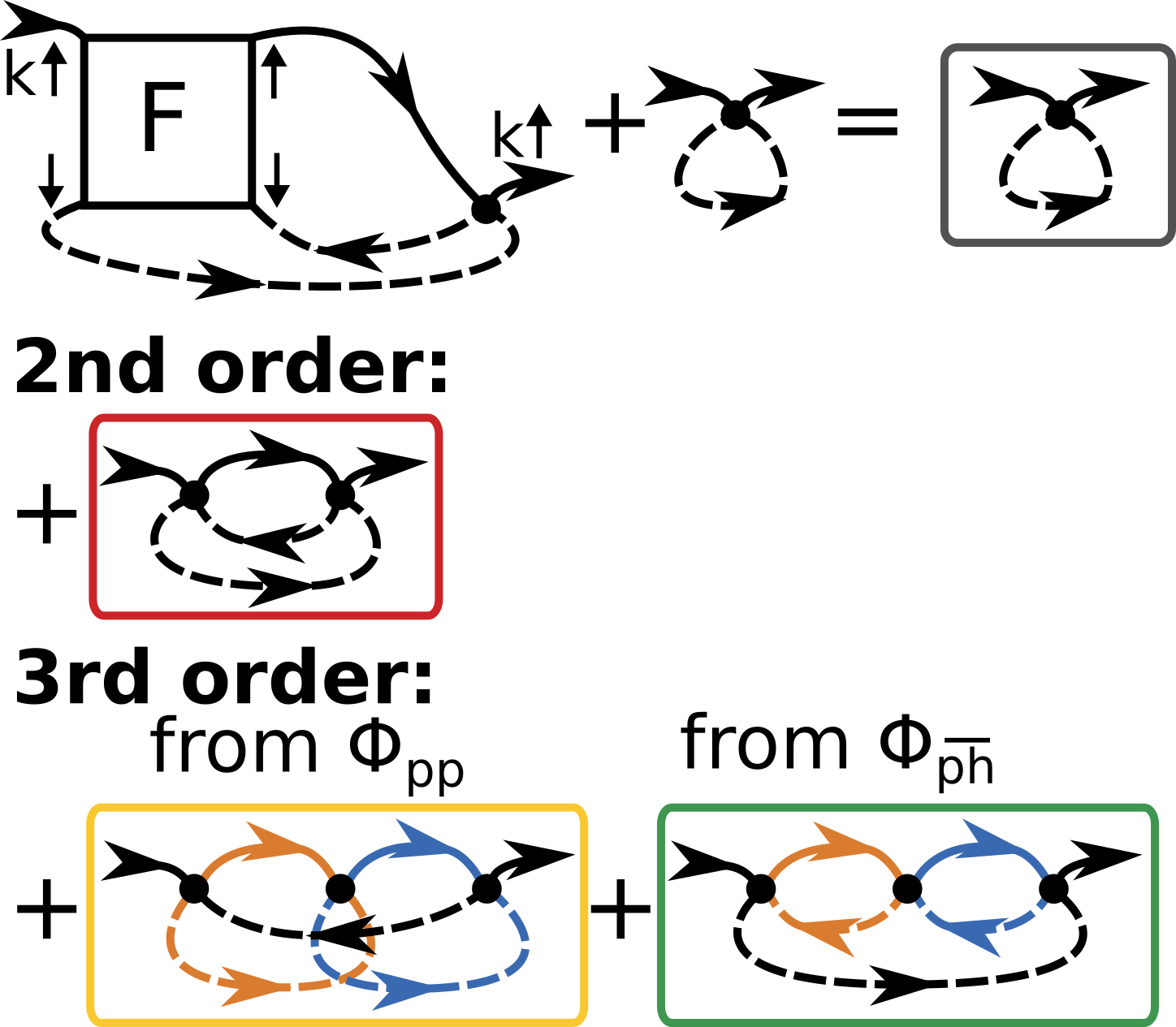}
    \caption{Lowest order diagrammatic contributions to the SDE self-energy flow, where solid (dashed) lines carry spin up (down) and the orange bubbles are projected to the blue ones. The colored boxes facilitate the comparison to the conventional flow shown in Fig.~\ref{fig:Perturbation_flow} (note that due to the restriction to bare Green's functions not all corresponding gray boxed diagrams are included). } 
    \label{fig:Perturbation_SDE}
\end{figure}

\begin{figure*}[ht]
    \centering
    \includegraphics[scale=0.4]{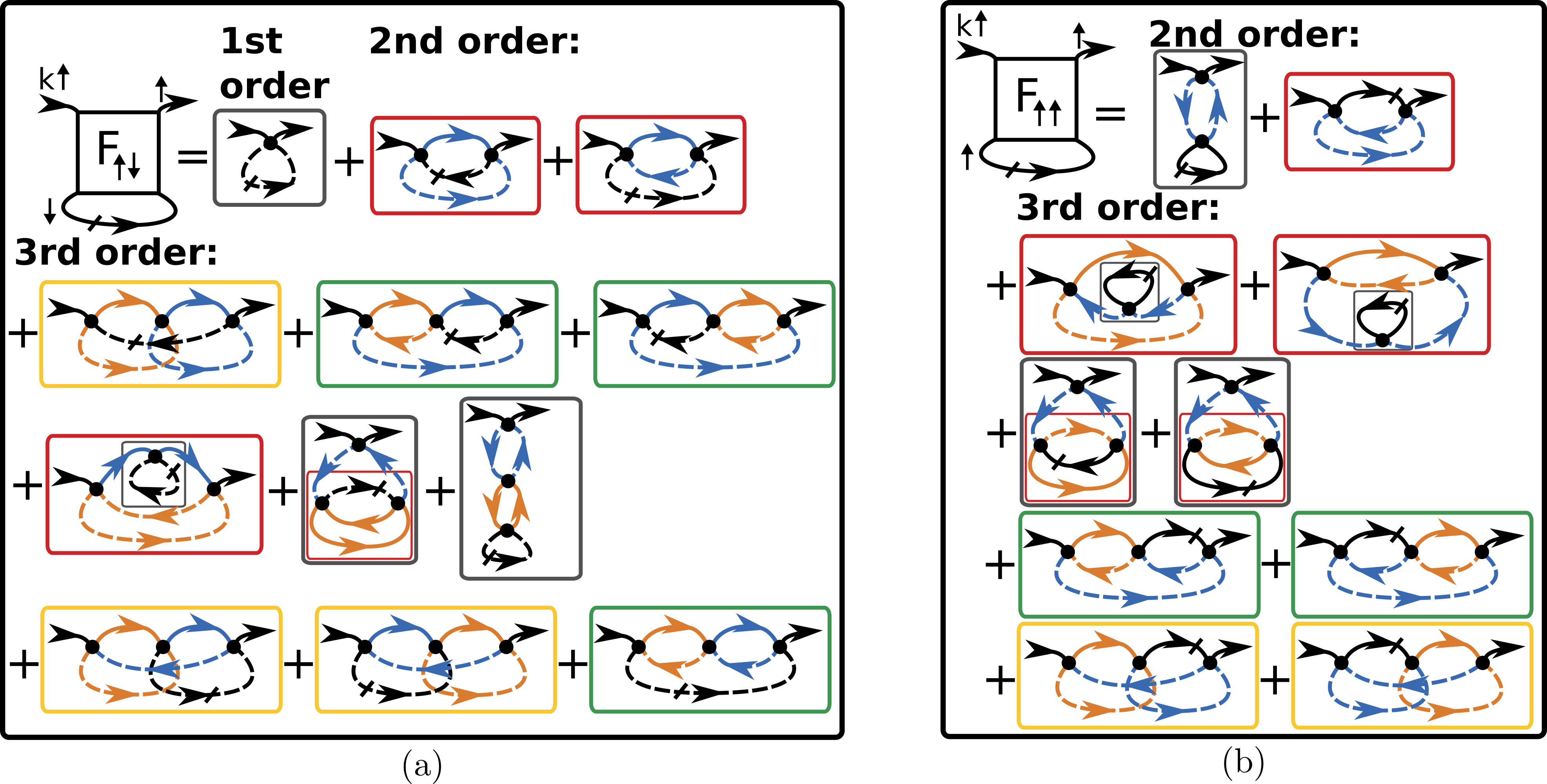}
    \caption{Lowest order diagrams for the conventional self-energy flow, with the (a) $F_{\uparrow\downarrow}$ and (b) $F_{\uparrow\uparrow}$ contribution. All propagators are bare ones (neglecting any self-energy correction). In order to facilitate the comparison with the SDE flow shown in Fig.~\ref{fig:Perturbation_SDE}, we group them in tadpole diagrams (gray boxes), equivalent 2nd order diagrams (red boxes), 3rd order diagrams which are formally equivalent to the 3rd order diagrams from $\Phi_{pp}$ ( $\Phi_{\overline{ph}}$) in the SDE flow (yellow/green boxes respectively) but partly affected by approximations due to the form-factor expansion and their projections in the different channels.}
    \label{fig:Perturbation_flow}
\end{figure*}

Our results do not contradict the findings of the parquet approximation \cite{Eckhardt2019}. There, for $U=2$ a pseudogap occurs at $1/T=26$ and a full gap at $1/T=30$. These temperatures are presently difficult to access within a multiloop fRG calculation, due to the required refinement of the momentum and frequency dependence of the two-particle vertex and the high number of loop orders needed for convergence.

\subsection{Difference of self-energy flow schemes in the TU-fRG}
\label{ssec:SEinTUfRG}

In this section we will discuss the origin of the different gap opening tendencies in the $1\ell$ and SDE flow and in particular address the question why the long-range AF correlations lead to a gap opening in the flow of the SDE, while its effect is weakened in the conventional $1\ell$ flow of the self-energy. 

In the multiloop expansion of the fRG, the two schemes are formally equivalent.  
Differences are introduced only by truncating the form-factor expansion, which prevents the full reconstruction of the SDE at loop convergence.
The results and discussion in Section \ref{ssec:mfRG} rule out the finite loop order as being responsible for the qualitatively different physical behavior between the conventional and the SDE flow. 

The difference between the self-energy results as obtained from the two flow schemes is hence to be attributed to the truncation of the form-factor expansion.
As discussed in Section~\ref{ssec:mfRG}, the approximations introduced by the projections between the different channel representations imply that the $1\ell$ and SDE flow are not equivalent any more. 
We first focus on the SDE~\eqref{eq:SDE} together with the corresponding flow Eq.~\eqref{eq:sigmasplit} and consider the lowest orders in the bare interaction $U$, see Fig.~\ref{fig:Perturbation_SDE}.
We decompose the vertex into the different channel contributions and for simplicity neglect the self-energy corrections in the propagators. The second order diagram is not associated to any specific channel and can be computed by using Fast-Fourier-transforms, see also Eq.~\eqref{eq:sigma_GGG}. 
For a better comparison to the conventional $1\ell$ flow, the different colors of the boxes indicate the topology while the ones of the Green's functions refers to the projections, i.e. the order in which the (truncated to $s$-wave) bubbles are inserted. In the present convention, the orange bubble is inserted into the blue one which is then closed by a single Green's function (black). We note that for the diagrams shown in Fig.~\ref{fig:Perturbation_SDE}, the restriction to $s$-wave does not introduce any approximation in the projection of the different channel representations.

\begin{figure}[b]
    \centering
    \includegraphics[scale=0.4]{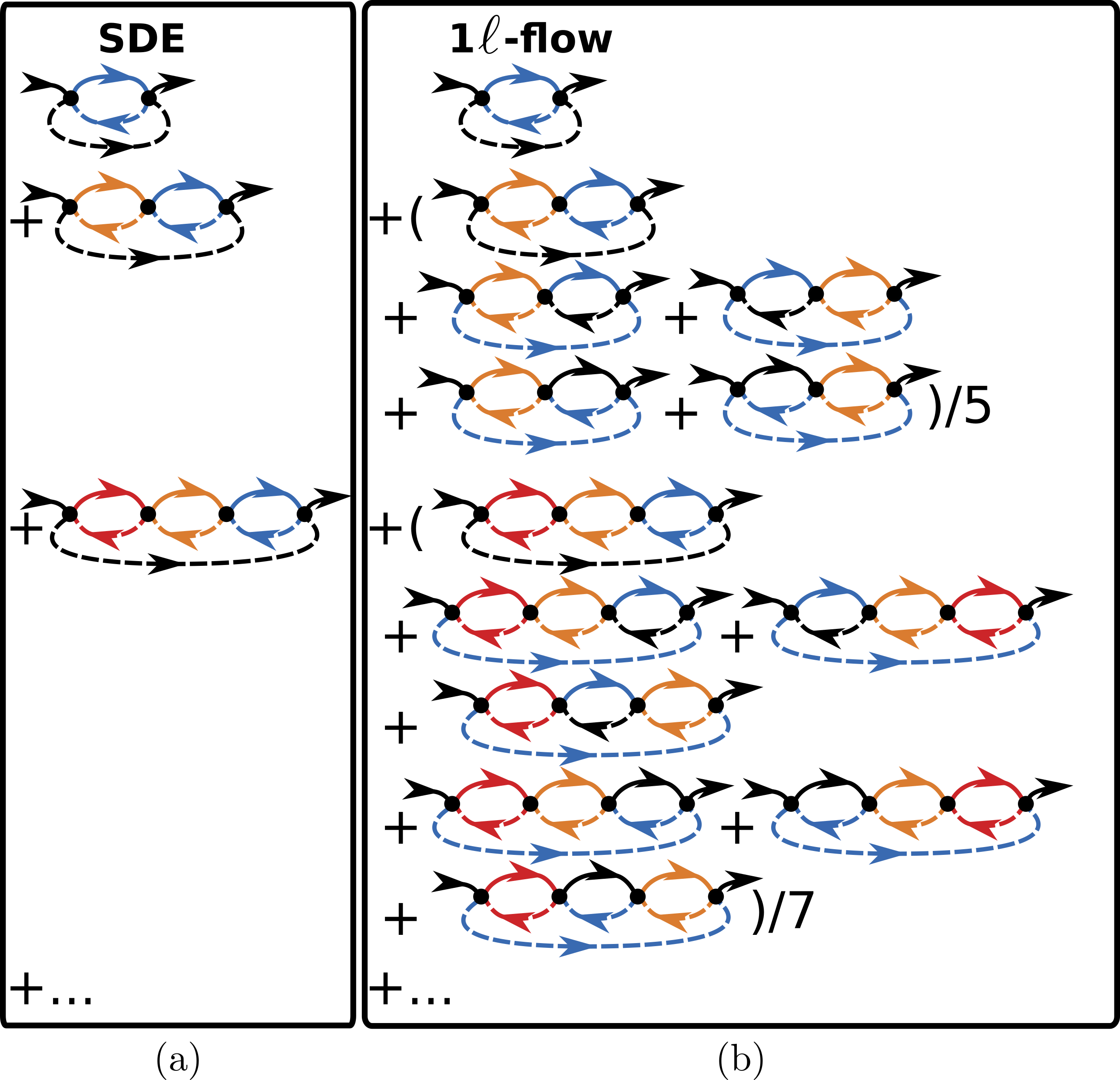}
    \caption{Comparison of the lowest order diagrams describing the contribution of the crossed particle-hole channel to the self-energy, (a) in the SDE and (b) in the conventional $1\ell$ flow. The results are shown in Figs. \ref{fig:Perturbation_xph_4th_7th} and \ref{fig:Gap_opening_vs_perturbation_order}.}
    \label{fig:Perturbation_xph_comparison}
\end{figure}

Now we compare these diagrams with those of the standard $1\ell$ flow shown in Fig.~\ref{fig:Perturbation_flow}. 
We insert the parquet decomposition of the two-particle vertex into the flow equation for the self-energy, being aware that in the $1\ell$ approximation this holds only up to second order, and group the different contributions according to their topology (indicated by the colored boxes), in analogy to Fig.~\ref{fig:Perturbation_SDE}. 
The gray box at first order (a) is the tadpole diagram, while the gray box at second order (b) is a self-energy correction thereof, which is included in the SDE-flow once the self-energy corrections in the Green's functions are accounted for (despite not being displayed in Fig.~\ref{fig:Perturbation_SDE}). The red ones at the second order include three contributions. These amount to exactly the same diagram in the SDE (see Fig.~\ref{fig:Perturbation_SDE}), as the blue bubble has no other contributions than $s$-wave. 
At third order, the red boxes with internal gray boxes are associated to self-energy corrections of the second order diagram, while the gray ones with and without internal red boxes correspond to self-energy corrections of the first order diagram. 
Most importantly, there are five different yellow and green boxes at third order which represent the contributions to the particle-particle and crossed particle-hole diagram in the SDE (see Fig.~\ref{fig:Perturbation_flow}). Here only a single one of each is correctly accounted for in $s$-wave. 
That the others are only correct in the infinite form factor limit, can be seen from the example of the first green diagram in the second line for $F_{\uparrow\downarrow}$: here the blue bubble gets non-$s$-wave contributions due to the insertion of the orange bubble. In the $s$-wave (or any few form factor) truncation, these are neglected. This explains why we obtain different results for the two self-energy flow schemes.

\begin{figure}[t]
    \centering
    \includegraphics[width=\cw]{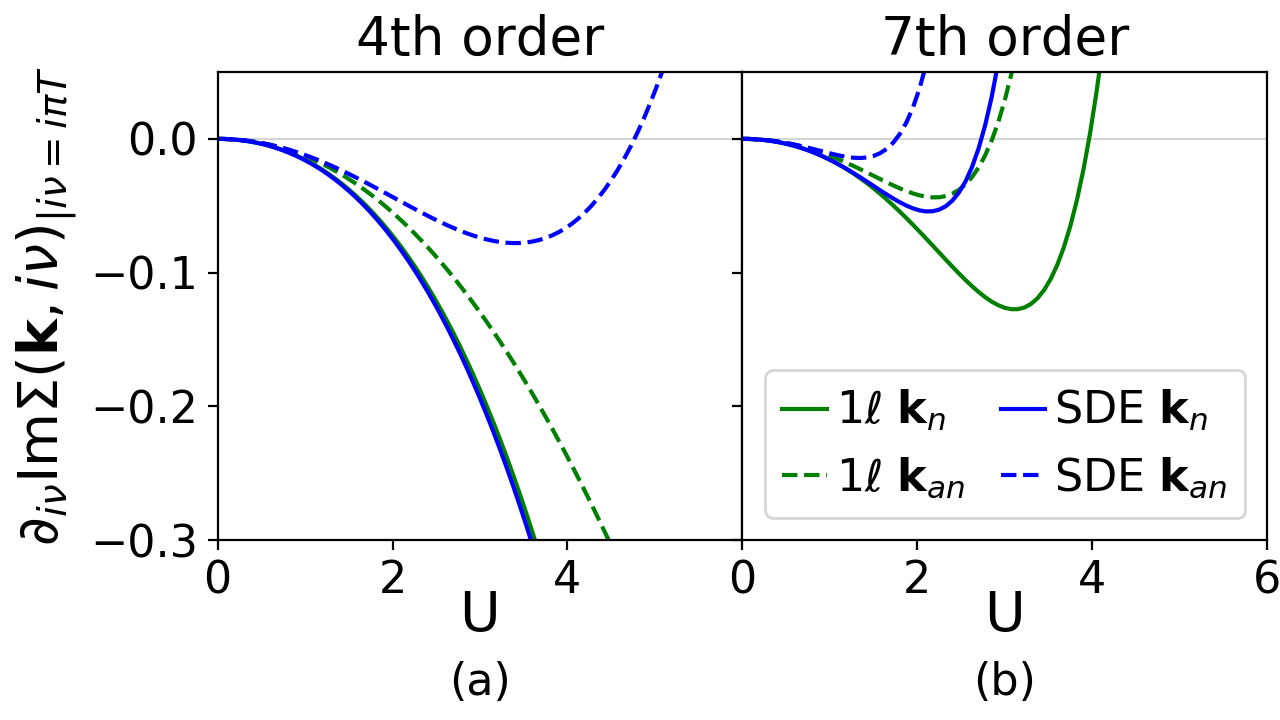}
    \caption{$\partial_{i\nu}\operatorname{Im}\Sigma({\bf k},i\nu)$ evaluated at $i\nu=i\pi T$ resulting from the contributions of the crossed particle-hole channel up to the (a) 4th and (b) 7th order in $U$, as illustrated in Fig.~\ref{fig:Perturbation_xph_comparison}. While at the second order no gap opens, starting from the 3rd both approaches lead to a gap opening at the nodal and antinodal point at some large value of $U$ (not shown). At the 4th order, the SDE flow opens a gap at the antinodal point first. At the 7th order one can already see that both flow schemes open a gap first at the antinodal point and then at the nodal one. Note that in the SDE flow both gaps open before the first gap opening in the conventional $1\ell$ flow sets in. For a study of the gap opening as a function of order in $U$ see Fig.~\ref{fig:Gap_opening_vs_perturbation_order}.}
    \label{fig:Perturbation_xph_4th_7th}
\end{figure}

\begin{figure}[b]
    \centering
    \includegraphics[width=\columnwidth]{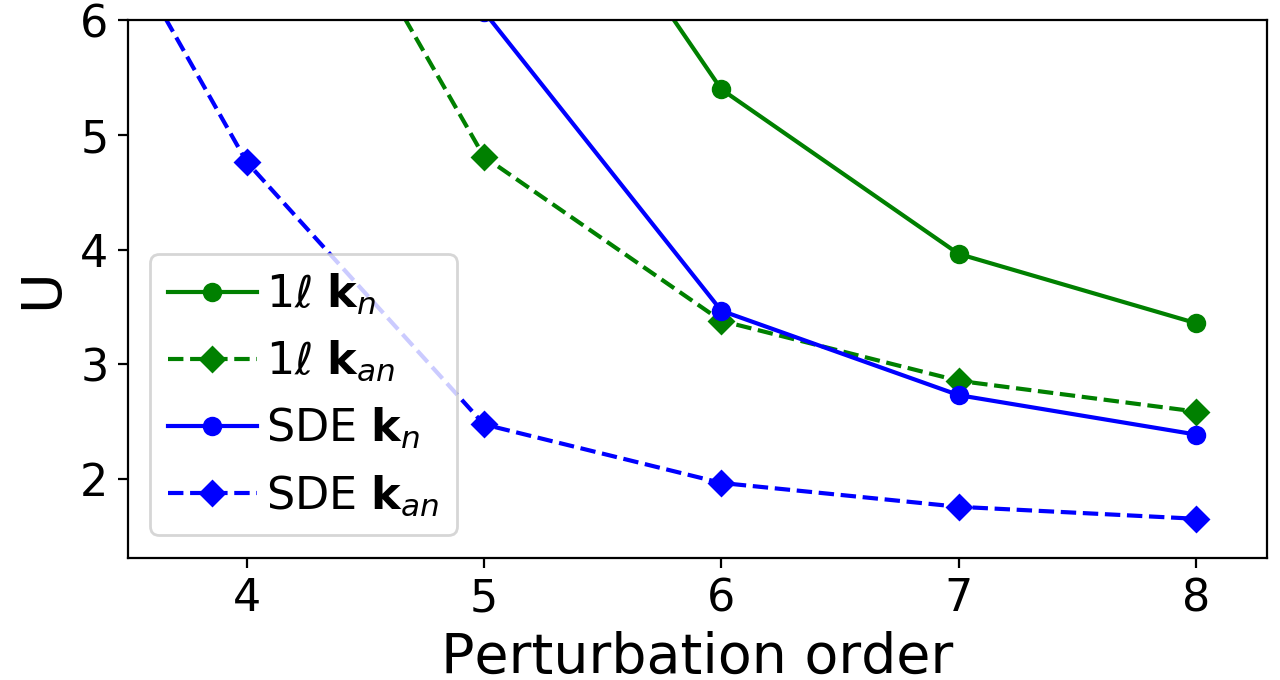}
    \caption{Gap opening as a function of order in $U$ at $1/T=10$ as extracted from $\partial_{i\nu}\operatorname{Im}\Sigma({\bf k},i\nu)$ evaluated at $i\nu=i\pi T$, resulting from the contributions of the crossed particle-hole channel only. With increasing order, the gap opening occurs at lower values of $U$. The gap at the antinodal point opens first, then the gap at the nodal point, consistently with the full calculation in Fig.~\ref{fig:Zfactor_flow_vs_SDE}. 
    While the qualitative behavior of the SDE and $1\ell$ flow is the same, the gaps in the former open first, followed by the ones in the latter.  
    Note that the pseudo-critical interaction is not captured in this approach as a divergence of the vertex occurs only at infinite order.
    }
    \label{fig:Gap_opening_vs_perturbation_order}
\end{figure}

In order to underline that the different form factor approximation of the crossed particle-hole channel is responsible for the flow-scheme-dependent gap opening tendencies, we analyze the lowest order contributions of the latter (e.g. the green boxes in Fig.~\ref{fig:Perturbation_SDE} represent the second order). 
The corresponding self-energy diagrams up to fourth order are shown in Fig.~\ref{fig:Perturbation_xph_comparison}, the extension to higher orders is straightforward.
The resulting $\partial_{i\nu}\operatorname{Im}\Sigma({\bf k},i\nu)$ is reported in Fig.~\ref{fig:Perturbation_xph_4th_7th} and \ref{fig:Gap_opening_vs_perturbation_order} for the 4th and 7th order in $U$.
The downturn of $\partial_{i\nu}\operatorname{Im}\Sigma({\bf k},i\nu)$ at small $U$ is due to the second order diagram in Fig.~\ref{fig:Perturbation_xph_comparison}. 
This is compensated by higher order contributions leading to a zero in $\partial_{i\nu}\operatorname{Im}\Sigma({\bf k},i\nu)$. The effect of the crossed particle-hole channel contributions is more pronounced at the antinodal point, see Fig.~\ref{fig:Perturbation_xph_4th_7th}. Concerning the two flow schemes, in the SDE flow the zero crossing is observed first at 4th order at roughly $U=4.8$, while all other crossings occur only at much larger values of $U$. For higher orders, the zero crossings shift to smaller interactions. 
At the 7th order both flows open a gap first at the antinodal point and then at the nodal one, consistently with the physical picture obtained by the fRG results of Fig.~ \ref{fig:Zfactor_flow_vs_SDE}. Note that in the SDE flow both gaps open before the first gap opening in the conventional $1\ell$ flow sets in.
This can seen also in the gap opening as a function of order in $U$, see Fig.~\ref{fig:Gap_opening_vs_perturbation_order}. With increasing order in $U$, the gap opening occurs at lower values of $U$.  
While the qualitative behavior of the SDE and $1\ell$ flow is the same, the gaps at the antinodal and then at the nodal point open first in the former, followed by the ones in the latter.

We note that in the fRG calculation the gap opening sets in at lower interactions due to the resummation of all orders. In fact, the perturbation theory does not capture the AF divergence.
Despite these limitations, the present analysis illustrates the order in which the gap opening occurs: first in SDE-flow at the antinodal and then nodal point; secondly, if we ignore the vertex divergency, in the $1\ell$ flow in the antinodal and last at the nodal point.

Finally, we emphasize that the two schemes yield different results only in a truncated form-factor expansion. A straightforward patching of all three momentum dependencies on the same grid involves no information loss in the projection from one channel to another. At the same time, the fine grid required to resolve the AF divergence in the magnetic channel, implies a huge numerical cost. 
Similarly, also in the TU-fRG scheme both schemes should converge to the same result for an increasing number of form factors. However, this number may be as large as the bosonic patching points.

\section{Conclusions and outlook}
\label{sec:conclusion}

We have presented numerical results for the electron self-energy of the 2D Hubbard on the square lattice at half filling and perfect nesting, using a forefront implementation of the TU-fRG. The main finding is that our refined prescription for the calculation of the self-energy permits to see the opening of a mildly anisotropic pseudogap at low temperatures and sufficient coupling strengths. Although foreshadowed in early $N$-patch fRG works \cite{Katanin2004,Rohe2005}, seeing the pseudogap opening had not been possible at the same degree of quantitative control with previous fRG approaches. The key insight of this paper on TU-fRG is that the truncation of the form-factor expansion for the fermionic momentum dependence of the interaction does affect the self-energy flow differently depending on how the latter is computed. Our new way via the Schwinger-Dyson equation reduces the truncation loss for the self-energy and thus allows one to observe the pseudogap opening.

While in this work we only claim qualitative accuracy, we have argued previously \cite{Hille2020} that the same (multiloop) fRG implementation also compares favorably with novel numerical solutions of the parquet equation (in the so-called parquet approximation), and most importantly, with determinant Quantum Monte Carlo, in situations where the sign-problem is not present. This emphasizes the high degree of numerical control of the method and that one can obtain not only qualitative information but also quantitatively correct results, at least for the one-band Hubbard model. 

We believe that fRG should be a welcome addition to the toolbox of theoretical methods that can compute spectral properties, as in general it has a high momentum resolution, decent flexibility regarding the model and system parameters, and an advantageous transparency as it allows one to identify the relevant interaction processes for a specific phenomenon. The next phase of research should now strive for improved numerical performance of the current fRG algorithm such that spectral and ordering properties of a wider class of correlated quantum materials can indeed be studied at a similar quantitative level.

\section{Acknowledgments}
The authors thank C. J. Eckhardt, A. Kauch, F. B. Kugler,  W. Metzner, T. Sch\"afer, A. Toschi, and D. Vilardi for valuable discussions. 
We acknowledge financial support from the Deutsche Forschungsgemeinschaft (DFG) through ZUK 63 and Projects No. AN 815/4-1 and No. AN 815/6-1. The authors also gratefully acknowledge the computing time granted through JARA on the supercomputer JURECA at Forschungszentrum J\"ulich \cite{jureca}.

\appendix

\section{Scale invariance in the multiloop 
interaction flow}
\label{app:scale_intflow}

In the interaction flow, the scale $\Lambda$ can be translated to the effective interaction \cite{Honerkamp2004}. 
This implies that at the scale $\Lambda$, the flow for a bare interaction $U$ corresponds to the final result of a flow with bare interaction $\Lambda^2U$.
We here prove this property to hold for all loop orders, specifically
\begin{subequations}
\label{eq:result}
\begin{align}
\label{eq:result_Sigma}
\Sigma_{\Lambda/l}(l^2U) &=l \Sigma_{\Lambda}(U)\\
V_{\Lambda/l}(l^2U)&=l^2 V_{\Lambda}(U)\;, \label{eq:result_V}
\end{align}
\end{subequations}
where the flow scale is indicated by the subscript and the bare interaction in the brackets. If $V_{\Lambda}(U)$ diverges at some $\Lambda<1$, we can set $l=\Lambda$ and find that the vertex would diverge exactly at $\Lambda=1$ for a bare interaction $\Lambda^2U$. 

If at every loop order
\begin{subequations}
\label{eq:condition}
\begin{align} 
    \label{eq:condition_Sigma}
    \dot{\Sigma}_{\Lambda/l}(l^2U)=l^2 \dot{\Sigma}_{\Lambda}(U)\\
    \label{eq:condition_V}
    \dot{V}_{\Lambda/l}(l^2U) = l^3 \dot{V}_{\Lambda}(U) \;.
\end{align}
\end{subequations}
is satisfied, we can show Eqs.~\eqref{eq:result} by using the following induction procedure. It is assumed that the integration in $\Lambda$ is done in descrete steps. At each step $n$, $\Lambda$ takes the value $\Lambda_n$ and we have to consider the equation 
\begin{align}
    V_{\Lambda_n}(U)&=V_{\Lambda_{n-1}}(U) + (\Lambda_n-\Lambda_{n-1}) \dot{V}_{\Lambda_n}(U) \;.
\end{align}
As base case 
we consider 
\begin{align} 
    V_{\Lambda_0/l}(l^2U)&=l^2V_{\Lambda_0}(U) =l^2U\;.
\end{align}
With the induction hypothesis~\eqref{eq:result_V} and the condition~\eqref{eq:condition_V} we can perform the induction step
\begin{align}
    V_{\Lambda_n/l}(l^2U)&=V_{\Lambda_{n-1}/l}(l^2U) + \left(\frac{\Lambda_n}{l}-\frac{\Lambda_{n-1}}{l}\right) \dot{V}_{\Lambda_n/l}(l^2U)\nonumber \\
    &= l^2 V_{\Lambda_{n-1}}(U) + l^2 (\Lambda_n-\Lambda_{n-1}) \dot{V}_{\Lambda_n}(U)\nonumber\\
    &= l^2 V_{\Lambda_n}(U) \;.
\end{align}
The same procedure can be performed for the self-energy with
\begin{align}
    \Sigma_{\Lambda_n}(U)&=\Sigma_{\Lambda_{n-1}}(U)+ (\Lambda_n-\Lambda_{n-1})  \dot{\Sigma}_{\Lambda_n}(U)
\end{align}
and the base case
\begin{align}
    \Sigma_{\Lambda_0/l}(l^2U) & = \Sigma_{\Lambda_0}(U) = 0\;.
\end{align}
Using the assumption~\eqref{eq:result_Sigma} and the condition~\eqref{eq:condition_Sigma},
the induction step yields
\begin{align}
    \Sigma_{\Lambda_n/l}(l^2U) &= \Sigma_{\Lambda_{n-1}/l}(l^2U) + \left(\frac{\Lambda_n}{l}-\frac{\Lambda_{n-1}}{l}\right) \dot{\Sigma}_{\Lambda/l}(l^2U)\nonumber \\ \label{eq:SE_induction_step}
    &= l \Sigma_{\Lambda_{n-1}}(U) + l (\Lambda_n-\Lambda_{n-1}) \dot{\Sigma}_{\Lambda}(U) \nonumber\\
    &= l\Sigma_{\Lambda_n}(U)\;.
\end{align}

For the proof of the conditions~\eqref{eq:condition} we have to use the Green's function and single-scale propagator, which in the interaction flow are defined as
\begin{subequations}
\begin{align} 
    G_{\Lambda}(U)  &=\Lambda\frac{1}{i\omega + \epsilon(k) -\Lambda \Sigma_{\Lambda}(U)} \label{eq:G_lam_SE_fb}&\\
    S_{\Lambda}(U) &= \frac{i\omega + \epsilon(k)}{(i\omega + \epsilon(k) -\Lambda \Sigma_{\Lambda}(U))^2} & \label{eq:S_lam_SE_fb}
\end{align}
\end{subequations}
respectively, and hence
\begin{subequations}
\begin{align}
    G_{\Lambda/l}(l^2U) &= \frac{\Lambda}{l} \frac{1}{i\omega + \epsilon(k) -\frac{\Lambda}{l} \Sigma_{\Lambda/l}(l^2U)} &\nonumber \\
    &= \frac{\Lambda}{l}\frac{1}{i\omega + \epsilon(k) -\Lambda \Sigma_{\Lambda}(U)} 
    =\frac{1}{l} G_{\Lambda}(U) &\label{eq:G_scaling_lam_SE_fb}\\
    S_{\Lambda/l}(l^2U) &= \frac{i\omega + \epsilon(k)}{(i\omega + \epsilon(k) - \frac{\Lambda}{l} \Sigma_{\Lambda/l}(l^2U))^2} \nonumber \\
    &= \frac{i\omega + \epsilon(k)}{(i\omega + \epsilon(k) - \Lambda \Sigma_{\Lambda}(U))^2} 
    =  S_{\Lambda}(U)\;, \label{eq:S_scaling_lam_SE_fb}
\end{align}
\end{subequations}
where we used the assumption~\eqref{eq:result_Sigma}. The single-scale propagator retains the same scaling property also after the Katanin substitution
\begin{align}
    S_{\Lambda}^K(U)&=\frac{i\omega + \epsilon(k) + \Lambda^2 \dot{\Sigma}_{\Lambda}(U)}{(i\omega + \epsilon(k) -\Lambda \Sigma_{\Lambda}(U))^2} \nonumber\\
    &= S_{\Lambda}(U) + G_{\Lambda}(U) \dot{\Sigma}_{\Lambda}(U) G_{\Lambda}(U)  
\end{align}
as we can use the condition~\eqref{eq:condition_Sigma} and the scaling property of the Green's function~\eqref{eq:G_scaling_lam_SE_fb} in 
\begin{align}
\hspace{-.1cm}S^K_{\Lambda/l}(l^2U) &= S_{\Lambda/l}(l^2U) \nonumber + G_{\Lambda/l}(l^2U) \dot{\Sigma}_{\Lambda/l}(l^2U) G_{\Lambda/l}(l^2U) \\
\hspace{-.1cm}    &= S_{\Lambda}(U) + \frac{1}{l} G_{\Lambda}(U) l^2 \dot{\Sigma}_{\Lambda}(U) \frac{1}{l} G_{\Lambda}(U) \nonumber\\
\hspace{-.1cm}    &= S^k_{\Lambda}(U)\;. \label{eq:SK_scaling}
\end{align}

For the proof of the conditions~\eqref{eq:condition}, we restrict ourselves to a single channel for simplicity (it holds analogously for their combination). In the $1\ell$ approximation, Eqs.~\eqref{eq:condition} are shown easily via
\begin{widetext}
\begin{subequations}
\begin{align}
   \dot{\Sigma}_{\Lambda/l}(l^2U) &= \SumInt
    S_{\Lambda_n/l}(l^2U)V_{\Lambda_{n-1}/l}(l^2U) 
    = l^2 \SumInt
    S_{\Lambda_n}V_{\Lambda_{n-1}}(U)
    = l^2 \dot{\Sigma}_{\Lambda}(U) \label{eq:proof_1l_Sigma}\\
    \dot{V}_{\Lambda_n/l}(l^2U)&=\SumInt V_{\Lambda_{n-1}/l}(l^2U)G_{\Lambda_n/l}(l^2U)S_{\Lambda_n/l}(l^2U)V_{\Lambda_{n-1}/l}(l^2U) \nonumber \\
    &=l^3 \SumInt V_{\Lambda_{n-1}}(U)G_{\Lambda_n}(U)S_{\Lambda_n}(U)V_{\Lambda_{n-1}}(U)  
    = l^3 \dot{V}_{\Lambda_n}(U) \label{eq:proof_1l_V}
 \;
\end{align}
\end{subequations}
\end{widetext}
which are satisfied both with and without self-energy feedback and also for the Katanin substitution replacing $S_{\Lambda}(U)$ by $S^K_{\Lambda}(U)$. For the m$\ell$ proof, we need to show in addition that if the equations at a specific $\Lambda_n$ are true for the loop order $\ell$, they are also true for $\ell+1$. For this induction we consider left, right and central diagrams: the left m$\ell$ correction is
\begin{align}
    \dot{\Phi}_{\eta,\Lambda}^{\textrm{left},\ell+2} = \dot{I}_{\eta,\Lambda}^{\ell+1} \circ \Pi_{\eta,\Lambda} \circ V_{\eta,\Lambda} \; , \label{eq:left_ml_corr_equation}
\end{align}
where $\dot{I}_{\eta,\Lambda}^{L}=\sum_{\eta'\neq \eta}\dot{\Phi}_{\eta',\Lambda}^{L}$ and $\eta$ denotes the channel and will be omitted in the following; the right m$\ell$ correction is related to Eq. \eqref{eq:left_ml_corr_equation} by an exchange of the position of $\dot{I}$ and $V$ which does not change the structure and hence the scaling property; the central m$\ell$ correction reads
\begin{align}
    \dot{\Phi}_{\Lambda}^{\textrm{central},\ell+2} = V_{\Lambda} \circ \Pi_{\Lambda} \circ \dot{I}_{\Lambda}^{\ell} \circ \Pi_{\Lambda} \circ V_{\Lambda} \; . \label{eq:central_ml_corr_equation}
\end{align}
The left, right and central contributions are added to the $1\ell$ vertex flow, which has been considered above and shown to satisfy the scaling property. Therefore the remaining equation to be proven flow step after flow step and loop order after loop order is 
\begin{widetext}
\begin{align}
    \dot{V}_{\Lambda_n}^{\ell}(U)=& \,\dot{I}_{\Lambda_n}^{\ell-1}(U) G_{\Lambda_n}(U) G_{\Lambda_n}(U) V_{\Lambda_{n-1}}(U) + V_{\Lambda_{n-1}}(U) G_{\Lambda_n}(U) G_{\Lambda_n}(U) \dot{I}_{\Lambda_n}^{\ell-1}(U)  \nonumber \\
    &+ V_{\Lambda_{n-1}}(U) G_{\Lambda_n}(U) G_{\Lambda_n}(U) \dot{I}_{\Lambda_n}^{\ell-2}(U) G_{\Lambda_n}(U) G_{\Lambda_n}(U) V_{\Lambda_{n-1}}(U)\;. \label{eq:equation_ml_V}
\end{align}
\end{widetext}
While for $\Lambda_n=\Lambda_0$, we get trivially for all loop orders
\begin{align}
    \dot{V}^{\ell}_{\Lambda_0/l}(l^2U)=0=\dot{V}^{\ell}_{\Lambda_0}(U) \;,
\end{align}
for $\Lambda_1$, we refer to the following equations for any flow step $\Lambda_n$. There, we have to consider two base cases as the multiloop correction according to Eq.~\eqref{eq:equation_ml_V} depends both on  $\ell-1$ and $\ell-2$. The first is actually the $\ell=1$ contribution~\eqref{eq:proof_1l_V} and the second the $\ell=2$ contribution
\begin{widetext}
\begin{align}
    \dot{V}^{\ell=2}_{\Lambda_n/l}(l^2U)&= \SumInt \underbrace{V^{\ell=1}_{\Lambda_n/l}(l^2U) S_{\Lambda_n/l}(l^2U) G_{\Lambda_n/l}(l^2U) V^{\ell=1}_{\Lambda_n/l}(l^2U)}_{\dot{I}_{\Lambda_n/l}^{\ell=1}(l^2U)}  G_{\Lambda_n/l}(l^2U) G_{\Lambda_n/l}(l^2U) V_{\Lambda_{n-1}/l}(l^2U)\nonumber \\
    &+ \textrm{(right} \sim \textrm{left)} + \textrm{(central}=0\textrm{ )} \nonumber\\
    &= l^3 \SumInt V^{\ell=1}_{\Lambda_n}(U) S_{\Lambda_n}(U) G_{\Lambda_n}(U) V^{\ell=1}_{\Lambda_n}(U) G_{\Lambda_n}(U) G_{\Lambda_n}(U) U 
    \hspace{1cm}= l^3 \dot{V}^{\ell=2}_{\Lambda_1}(U) \;. \label{eq:base2_ml_V_fin}
\end{align}
\end{widetext}
Here, we used Eq.~\eqref{eq:condition_V} for $V^{\ell=1}_{\Lambda_n/l}(l^2U)$ and also for $V_{\Lambda_{n-1}/l}(l^2U)$. The latter is only true if we apply the induction proof for each $\Lambda_n$ independently (and in increasing order of $n$). We note that if the condition~\eqref{eq:equation_ml_V} is true, then
\begin{align}
    \dot{I}^{\ell-1}_{\Lambda_n/l}(l^2U)&=\nonumber\\& V^{\ell-1}_{\Lambda_n/l}(l^2U) S_{\Lambda_n/l}(l^2U) G_{\Lambda_n/l}(l^2U) V^{\ell-1}_{\Lambda_n/l}(l^2U) \\
    \dot{I}^{\ell-2}_{\Lambda_n/l}(l^2U)&=\nonumber\\&V^{\ell-2}_{\Lambda_n/l}(l^2U)
    S_{\Lambda_n/l}(l^2U) G_{\Lambda_n/l}(l^2U) V^{\ell-2}_{\Lambda_n/l}(l^2U) \,.
\end{align}
We will use these equations in the induction step
\begin{widetext}
\begin{align}
    \dot{V}_{\Lambda_n/l}^{\ell}(l^2U) =&\, \dot{I}^{\ell-1}_{\Lambda_n/l}(l^2U) G_{\Lambda_n/l}(l^2U) G_{\Lambda_n/l}(l^2U) V_{\Lambda_{n-1}/l}(l^2U)  + \textrm{(right} \sim \textrm{left)}\nonumber\\
    & + V_{\Lambda_{n-1}/l}(l^2U) G_{\Lambda_n/l}(l^2U) G_{\Lambda_n/l}(l^2U) \dot{I}_{\Lambda_n/l}^{\ell-2}(l^2U) 
    G_{\Lambda_n/l}(l^2U) G_{\Lambda_n/l}(l^2U) V_{\Lambda_{n-1}/l}(l^2U) \nonumber\\ \label{eq:mfRG_induction_step}
    =& \,l^3 \dot{I}^{\ell-1}_{\Lambda_n}(U) \frac{1}{l} G_{\Lambda_n}(U) \frac{1}{l} G_{\Lambda_n}(U) l^2 V_{\Lambda_{n-1}}(U)  + \textrm{(right} \sim \textrm{left)} \nonumber \\
    & + l^2 V_{\Lambda_{n-1}}(U) \frac{1}{l} G_{\Lambda_n}(U) \frac{1}{l} G_{\Lambda_n}(U) l^3 \dot{I}_{\Lambda_n}^{\ell-2}(U) \frac{1}{l} G_{\Lambda_n}(U) \frac{1}{l} G_{\Lambda_n}(U) l^2 V_{\Lambda_{n-1}}(U) \nonumber\\
    =& \,l^3 \dot{V}^{\ell}_{\Lambda_n}(U) \;,
\end{align}
\end{widetext}
which proves the scaling property of the vertex flow.

Finally we turn to the two 
multiloop corrections of the self-energy, for which we simply have to show Eq.~\eqref{eq:condition_Sigma} in order to repeat the procedure as outlined in Eq.~\eqref{eq:SE_induction_step}. The first multiloop correction is 
\begin{align}
    \dot{\Sigma}^{(1)}_{\Lambda_n}(U) &= -\SumInt G_{\Lambda_n}(U) \left[ 2\dot{V}_{\Lambda_n}(U) - \dot{V}_{\Lambda_n}(U)\right] \;,
\end{align}
satisfying the condition of the scale property
\begin{align}
    \dot{\Sigma}^{(1)}_{\Lambda_n/l}(l^2U) &= -\SumInt G_{\Lambda_n/l}(l^2U) \nonumber \\
    &\hspace{1cm}\times\left[ 2\dot{V}_{\Lambda_n/l}(l^2U) - \dot{V}_{\Lambda_n/l}(l^2U) \right] \nonumber \\
    &= -\SumInt \frac{1}{l} G_{\Lambda_n}(U) \left[ 2 l^3 \dot{V}_{\Lambda_n}(U) - l^3 \dot{V}_{\Lambda_n}(U) \right]\nonumber \\
    &= l^2 \dot{\Sigma}^{(1)}_{\Lambda_n}(U) \,.
\end{align}
The second correction reads
\begin{align}
    \dot{\Sigma}^{(2)}_{\Lambda_n}(U) &= -\SumInt \delta S_{\Lambda_n}(U) \Big[2 V_{\Lambda_n}(U) - V_{\Lambda_n}(U) \Big]\;, 
\end{align}
where $\delta S_{\Lambda_n}(U)=G_{\Lambda_n}(U)\dot{\Sigma}^{(1)}_{\Lambda_n}(U)G_{\Lambda_n}(U)$ and also satisfies the condition
\begin{align}
    \dot{\Sigma}^{(2)}_{\Lambda_n/l}(l^2U) &= -\SumInt \delta S_{\Lambda_n/l}(l^2U)\nonumber \\
    &\hspace{1cm}\times \left[2 V_{\Lambda_n/l}(l^2U) - V_{\Lambda_n/l}(l^2U) \right] \nonumber \\
    &= -\SumInt \delta S_{\Lambda_n}(U) \left[ 2 l^2 V_{\Lambda_n}(U) - l^2 V_{\Lambda_n}(U) \right] \nonumber \\
    &= l^2 \dot{\Sigma}_{\Lambda_n}^{(2)}(U) \,,
\end{align}
where we used that $\delta S_{\Lambda_n}(U)$ scales as the single-scale propagator (with Katanin substitution) in Eq.~\eqref{eq:SK_scaling}. 

In addition, we verified also numerically that the scaling property is satisfied at any loop order.

\section{Importance of the crossed particle-hole contribution}
\label{app:xph}

Here we discuss the role of the crossed particle-hole channel $\Phi_{\overline{ph}}$ for the gap opening (see also Refs.~\cite{Vilk1997,Abanov2003} and~\cite{Montiel2017}). This channel translates directly to the magnetic channel, which dominates the physics at half filling and in a $1\ell$ fRG diverges at a pseudo-critical interaction. Let us therefore focus on
\begin{align}
    \Sigma_{\overline{ph}}({\bf k},i\nu) =& -\sum_{{\bf k'}i\nu'} \sum_{m}f^*_m({\bf k}) f_{0}({\bf k}) 4\pi^2 U \nonumber \\
    & \times \sum_{i\nu''} \sum_{n}
    \Phi_{\overline{ph}}({\bf k'}-{\bf k},m,n,i\nu'-i\nu,i\nu,i\nu'') \nonumber \\
    & \times\Pi_{ph}({\bf k'}-{\bf k},n,0,i\nu'-i\nu,i\nu'') 
    G({\bf k'},i\nu')  \label{eq:full_sigma_xph} \;
\end{align}
and neglect all other contributions to 
\begin{align}
    \Sigma({\bf k},i\nu) =& \Sigma_{G}(k,i\nu)   
     + \Sigma_{GGG}(k,i\nu) 
     + \Sigma_{pp}(k,i\nu)  \nonumber \\
     &+ \Sigma_{ph}(k,i\nu) 
     + \Sigma_{\overline{ph}}(k,i\nu) \label{eq:full_sigma_split}\;.
\end{align}
Close to the divergence, $\Phi_{\overline{ph}}$ exhibits a very strong $s$-wave component such that we can neglect other form-factor contributions and approximate Eq.~\eqref{eq:full_sigma_xph} by
\begin{align}
    \Sigma_{\overline{ph}}({\bf k},i\nu) \approx& -\sum_{{\bf k'}i\nu'}  U \nonumber \\    & 
     \times\sum_{i\nu''} 
    \Phi_{\overline{ph}}({\bf k'}-{\bf k},0,0,i\nu'-i\nu,i\nu,i\nu'') \nonumber \\
    & \times\Pi_{ph}({\bf k'}-{\bf k},0,0,i\nu'-i\nu,i\nu'') 
    G({\bf k'},i\nu')  \label{eq:full_sigma_xph_ff} \;.
\end{align}
Using the high frequency asymptotics of the two-particle vertex \cite{Wentzell2016} we obtain
\begin{align}
    \Sigma_{\overline{ph}}({\bf k},i\nu) \approx& -\sum_{{\bf k'}i\nu'}  \Big[  \mathcal{K}_{2,\overline{ph}}({\bf k'}-{\bf k},0,i\nu'-i\nu,i\nu)  \nonumber \\
    &+\mathcal{K}_{1,\overline{ph}}({\bf k'}-{\bf k},i\nu'-i\nu) \Big]  G({\bf k'},i\nu')  \label{eq:full_sigma_xph_asympt} \;.   
\end{align}
Here, $\mathcal{K}_{1,\overline{ph}}$ which is proportional to the crossed-particle hole or to the (negative) magnetic susceptibility, yields the strongest contribution and can be approximated by 
\begin{align}
    \mathcal{K}_{1,\overline{ph}}({\bf k'}-{\bf k},i\nu'-i\nu)& \nonumber \\
     & \hspace{-2.5cm} \approx\delta_{{\bf k'}-{\bf k}=(\pi,\pi)}\delta_{i\nu'-i\nu=0}\mathcal{K}_{1,\overline{ph}}((\pi,\pi),0) \nonumber \\,
        & \hspace{-2.5cm} \approx -2\delta_{{\bf k'}-{\bf k}=(\pi,\pi)}\delta_{i\nu'-i\nu=0}\chi_{\mathrm{AF}} \;,
\end{align}
leading to the following expression for the self-energy
\begin{align} 
    \Sigma_{\overline{ph}}({\bf k},i\nu) &\approx 2\chi_{\mathrm{AF}} G({\bf k}+(\pi,\pi),i\nu)  \label{eq:full_sigma_xph_delta} \\
    &\approx 2\chi_{\mathrm{AF}} \frac{1}{i\nu + \epsilon_{{\bf k}+(\pi,\pi)}} \;. \label{YRZsigma}
\end{align}

\begin{figure}[t]
    \centering
    \includegraphics[width=\columnwidth]{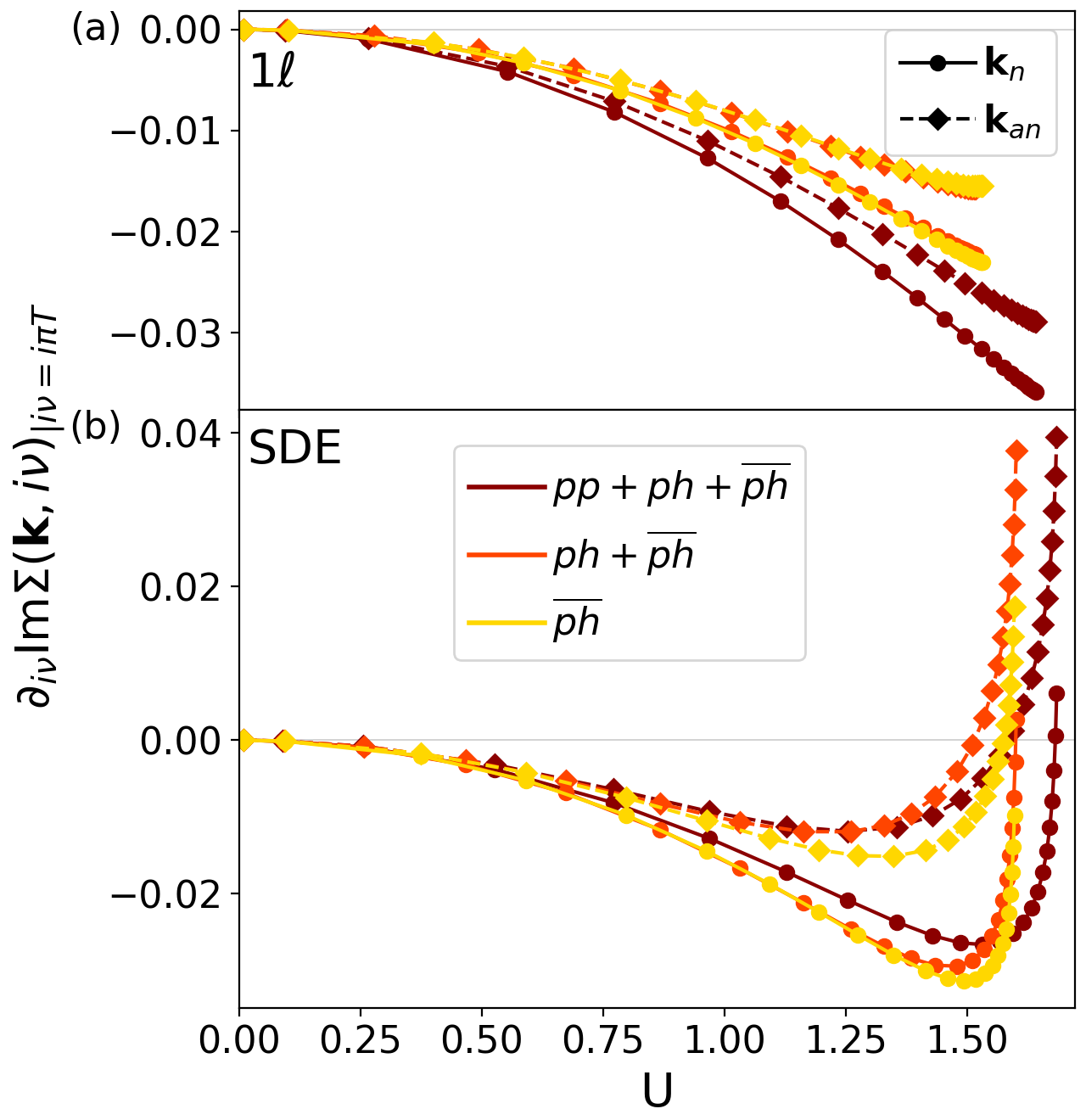}
    \caption{$\partial_{i\nu}\operatorname{Im}\Sigma({\bf k},i\nu)$ evaluated at $i\nu=i\pi T$ as a function of the flowing interaction $U$ for $1/T=10$ and different channel approximations, with both (a) the conventional fRG and (b) the SDE-approach for the self-energy flow. Neglecting the $pp$- and $ph$-channels does not qualitatively affect the appearance of the gap opening, except for the reduction of the critical scale which eventually prevents its observation at $k_n$.}
    \label{fig:Zfactor_channels}
\end{figure}

For momenta on the Fermi surface $\epsilon_{{\bf k}+(\pi,\pi)}=0$, the crossed particle-hole contribution to the imaginary part of the self-energy is $-2\chi_{\mathrm{AF}}/(\pi T)$ for the first and $-2\chi_{\mathrm{AF}}/(3\pi T)$ for the second Matsubara frequency. For momenta on the Fermi surface we thus obtain $\partial_{i\nu}\operatorname{Im}\Sigma({\bf k},i\nu)=2\chi_{\mathrm{AF}}/(3\pi^2T^2)>0$. For comparison we estimate $\operatorname{Im}\Sigma({\bf k},i\nu=i\pi T)=-2\pi T \chi_{\mathrm{AF}}/(\pi^2 T^2+16)$ and $\operatorname{Im}\Sigma({\bf k},i\nu=i3\pi T)=-6\pi T \chi_{\mathrm{AF}}/(9\pi^2 T^2+16)$ for the momenta ${\bf k}=(0,0)$ and ${\bf k}=(\pi,\pi)$, with $\epsilon_{(0,0)}=-\epsilon_{(\pi,\pi)}=- 4$. In this simplified analysis, $\partial_{i\nu}\operatorname{Im}\Sigma$ at these momenta is negative for
$T>\frac{4}{\sqrt{3}\pi}=0.735$ and therefore presents no gap opening for higher temperatures. 
 
Note that the self-energy in Eq. \eqref{YRZsigma} basically coincides with the phenomenological ansatz of Eq. 7 in \cite{Yang2006}. This indicates that the phenomenology arising from the numerical study here may indeed be useful to explain pseudogap features in correlated materials like high-temperature superconductors.

If the crossed particle-hole channel drives the gap opening, suppressing the contribution of the other channels should not change qualitatively the results, see Fig.~\ref{fig:Zfactor_channels}. Setting the particle-particle channel to zero (orange) or setting both the direct particle-hole and particle-particle channel to zero does not open a gap in the $1\ell$ scheme and preserves the gap at the antinodal point in the SDE scheme. At the nodal point, where the gap opening is unstable with respect to the momentum patching points and the loop order, no gap occurs when only the crossed particle-hole channel is taken into account.

\bibliography{bibliography}

\end{document}